\documentclass{ws-rv961x669}
\usepackage{graphicx}
\usepackage[utf8]{inputenc}
\usepackage{hyperref}
\usepackage{color}
\usepackage{slashed}
\usepackage{hyperref}
\usepackage[square]{ws-rv-van}
\usepackage{bm}
\def\simg{{\ \lower-1.2pt\vbox{\hbox{\rlap{$>$}\lower6pt\vbox{\hbox{$\sim$}}}}\ }}
\def\lsi{\raise0.3ex\hbox{$<$\kern-0.75em\raise-1.1ex\hbox{$\sim$}}}
\def\gsi{\raise0.3ex\hbox{$>$\kern-0.75em\raise-1.1ex\hbox{$\sim$}}}
\newcommand{\lsim}{\mathop{\lsi}}
\newcommand{\gsim}{\mathop{\gsi}}
\newcommand{\bsl}[1]{\,\slash\!\!\!\!{#1}\,}

\newcommand{\tr}{{\rm Tr\,}}
\usepackage{slashed}
\def\cropnote{\relax}

\begin{document}

\chapter*{Status of rates and rate equations for thermal leptogenesis \label{Ch4}}

\author[
]{S.~Biondini$^*$\footnote{Corresponding Author.}, D.~B\"odeker$^\dagger$, N.~Brambilla$^\ddagger$, M.~Garny$^\ddagger$, J.~Ghiglieri$^\S$, A.~Hohenegger$^\P$, M.~Laine$^*$, S.~Mendizabal$^\parallel$, P.~Millington$^{\ast\ast}$, A.~Salvio$^\S$, A.~Vairo$^\ddagger$}

\address{$^*$AEC, Institute for Theoretical Physics, University of Bern, \\ Sidlerstrasse 5, CH-3012 Bern, Switzerland \\[3pt]
$^\dagger$Fakult\"at f\"ur Physik, Universit\"at Bielefeld, 
\\
33501 Bielefeld, Germany \\[3pt]
$^\ddagger$Physik-Department, Technische Universit\"{a}t M\"{u}nchen,
\\
James-Franck-Str. 1, 85748 Garching, Germany  \\[3pt]
$^\S$CERN, Theoretical Physics Department, 
\\Geneva, Switzerland \\[3pt]
$^\P$Faculty of Science and Technology, University of Stavanger, 
\\
N-4036 Stavanger, Norway   \\[3pt] 
$^\parallel$Department of Physics, Universidad T\'ecnica Federico Santa Mar\'ia,
\\ Casilla 110-V, Valpara\'iso, Chile \\[3pt]
$^{\ast\ast}$School of Physics and Astronomy, University of Nottingham,\\
Nottingham NG7 2RD, UK
}

\begin{abstract}
\textbf{Abstract}: In many realizations of leptogenesis, heavy right-handed neutrinos play the main role in the generation of an imbalance between matter and antimatter in the early Universe. Hence, it is relevant to address quantitatively their dynamics in a hot and dense environment by taking into account the various thermal aspects of the problem at hand.  The strong washout regime offers an interesting framework to carry out calculations systematically and reduce theoretical uncertainties. Indeed, any matter-antimatter asymmetry generated when the temperature of the hot plasma $T$ exceeds the right-handed neutrino mass scale $M$ is efficiently erased, and one can focus on the temperature window $T \ll M$. We review recent progresses in the thermal field theoretic derivation of the key ingredients for the leptogenesis mechanism: the right-handed neutrino production rate, the CP asymmetry in the heavy-neutrino decays and the washout rates. The derivation of evolution equations for the heavy-neutrino and lepton-asymmetry number densities, their rigorous formulation and applicability are also discussed.
\end{abstract}

\body
\tableofcontents

\section{Introduction}\label{Ch4:sec1}
The explanation of the observed baryon asymmetry of the Universe is
an interesting and challenging endeavor for both cosmology and particle physics. 
Since any primordial imbalance between particles and antiparticles in the early Universe 
has likely been washed out after the inflationary epoch, a dynamical generation of the baryon asymmetry after the reheating phase appears favoured. Such dynamical generation in the context of quantum field theory is called baryogenesis.

One of the most attractive frameworks for baryogenesis is
via leptogenesis~\cite{Fukugita:1986hr}. In its original formulation, leptogenesis
demands a modest extension of the Standard Model (SM): namely, the
addition of right-handed (RH) neutrinos with large Majorana masses far
above the electroweak scale. The RH (or sterile) neutrinos are
singlets under the SM gauge group, whereas they are minimally coupled to the
SM particles via complex Yukawa couplings. The latter provide an additional
source of CP violation with respect to the one already present in the quark sector of the SM. 
The Lagrangian of the model, when expressing the RH neutrinos in a basis where the mass matrix is diagonal and with Majorana fields, reads\cite{Fukugita:1986hr}
\begin{equation}
\mathcal{L}=\mathcal{L}_{\hbox{\tiny SM}} 
+ \frac{1}{2} \,\bar{N}_{k} \,i \slashed{\partial}  \, N_{k}  - \frac{M_k}{2} \,\bar{N}_{k}N_{k} 
-   \left( \lambda_{\alpha k}\, \bar{\ell}_\alpha \phi^c N_k + \text{h.c.} \right) \, ,
\label{Lag1_full}
\end{equation} 
where $\mathcal{L}_{\hbox{\tiny SM}} $ is the SM Lagrangian, $N_k$ are Majorana spinors ($N_k=N_k^c$, $N_k^c$ being the charge-conjugate spinor of $N_k$) 
with mass $M_k$ and $k=1,2,3$ is the mass-eigenstate index.\footnote{There exist other formulations of the very same Lagrangian where the field $N_k$ is understood as a two-component spinor, or alternatively, as a four-component spinor where only the right-handed entries are non-zero (see, e.g.\cite{Drewes:2013gca}). They differ necessarily in the normalization of the kinetic terms.} In addition, $\ell_{\alpha} = (\nu_{L\alpha},e_{L\alpha})^T$ is the SM left-handed lepton doublet with $\alpha=e,\mu,\tau$, $\phi^c= \epsilon \, \phi^*$ ($\epsilon_{12}=-1$) is the isospin-conjugate of the Higgs doublet, $\lambda_{\alpha k}$ are complex Yukawa couplings. We shall adopt the following notation for the Yukawa couplings when summing over active lepton flavor: $\lambda^*_i \lambda_k \equiv \sum_\alpha \lambda^*_{\alpha i} \lambda_{\alpha k}$ (for $k=i$, $|\lambda_k|^2 \equiv \sum_\alpha \lambda^*_{\alpha k} \lambda_{\alpha k}$).  
    
In the standard picture, the RH neutrinos are produced by thermal scatterings in the early Universe and then decay out of equilibrium either into SM leptons or antileptons in different
amounts due to the CP-violating phases. Such an asymmetry in the
lepton sector is then partially converted into a baryon asymmetry by sphaleron processes in the SM\cite{Kuzmin:1985mm}. 

Majorana neutrinos decay in a hot medium, namely the Universe in its early stages.
Interactions with the medium modify the neutrino dynamics, through their production rate, masses and CP asymmetries, 
and affect the thermodynamic evolution of the lepton asymmetry. In order to take into account various thermal effects in leptogenesis properly, one has to consider the framework of quantum field theory at finite temperature. In so doing, all the ingredients in the analysis can be cast on a sound
theoretical footing. However, the derivation of
observables at finite temperature poses both conceptual and technical challenges. The purpose of the present paper is to review and report the recent results and developments on this subject. 

We shall discuss thermal leptogenesis in the so-called strong washout regime. In the literature, the strong washout scenario is usually defined in terms of the \textit{decay parameter} for the $k$-th RH neutrino, $K_k=\Gamma_k/ H$\cite{Buchmuller:2004nz,Davidson:2008bu}, which is the ratio between the neutrino decay rate and the Hubble rate. The former is taken at $T=0$ and the latter is evaluated at $T = M_k$.  For $K_k$ larger than one, the strong washout gets realized, enforcing decays and inverse decays to thermalize rapidly at $T \sim M_k$. Therefore, any initial lepton asymmetry possibly present before the onset of leptogenesis is erased\cite{Buchmuller:2004nz,Davidson:2008bu} and the neutrino dynamics is close to equilibrium. The decay parameter can in turn be related to active neutrino mass parameters\cite{Buchmuller:1996pa,Davidson:2008bu}, $K_k= \tilde{m}_k/m_{*}$, where $\tilde{m}_k=|\lambda_k|^2 v^2/(2M_k)$ (effective neutrino masses), $m_{*} \simeq 1.1 \times 10^{-3}$ eV (equilibrium neutrino mass) and $v=246$ GeV is the vacuum expectation value of the SM Higgs field. By assigning to the effective neutrino masses either the solar or atmospheric neutrino mass scale from the available neutrino mixing and oscillations data \cite{Fogli:2011qn,Blanchet:2012bk}, the decay parameter is estimated to be $\mathcal{O}(K)\sim 10$ -- $50$. It is then conceivable that the strong washout regime was established in the early Universe. In this case, the dependence on the initial conditions is practically absent. Leptogenesis may then be highly predictive, and we can focus on deriving relevant observables for temperatures $T \ll M$.

The outline of the chapter is as follows. In \sref{basics_lep}, we provide an introduction to thermal leptogenesis and the rates involving RH neutrinos. In \sref{Ch4:sec2}, the neutrino production rate is expressed in terms of a neutrino thermal width and we discuss how in-vacuum and thermal corrections can be derived. Both the non-relativistic regime and the more challenging relativistic regime are addressed. Then we introduce the CP asymmetry in RH neutrino decays in \sref{Ch4:sec3}. Thermal corrections to the CP asymmetry in the non-relativistic regime are reviewed, whose derivation are based on effective field theory (EFT) techniques. Moreover, their connection with the lepton asymmetry is discussed in the framework of non-equilibrium field theory. In \sref{Ch4:sec4}, the rate equations governing the evolution of the heavy-neutrino and lepton-asymmetry number densities are described. In particular, we focus on the applicability of the Boltzmann equations for leptogenesis and on different approaches to their generalization in a quantum field theory framework.  Conclusions and outlook are found in \sref{Ch4:sec5}.

\section{Basics of thermal leptogenesis}
\label{basics_lep}

The sections that follow review the recent progress made in the calculation of the various CP-even and -odd rates needed to obtain quantitatively accurate estimates of the final asymmetry in thermal leptogenesis. In order to motivate these discussions, we first outline the basics of thermal leptogenesis. Whilst the rest of this chapter will focus on first-principles and finite-temperature field-theoretic treatments, we will concentrate here on semi-classical treatments of leptogenesis, which involve supplementing systems of Boltzmann equations with the (zero-temperature) field-theoretic ingredients needed to capture the source(s) of CP violation. For more comprehensive overviews of the fundamentals of leptogenesis, see~Refs.~\cite{Pilaftsis:1998pd,Buchmuller:2004nz,Davidson:2008bu,Blanchet:2012bk,Fong:2013wr}.

The simplest scenario of thermal leptogenesis is realised when the heavy-neutrino mass spectrum is hierarchical, i.e.~$M_1\ll M_2<M_3$, and the decay rate of the lightest RH neutrino ($N_1$) is larger than the Hubble rate, i.e.~$\Gamma_1\gg H$. The dominant production of lepton asymmetry then occurs through the out-of-equilibrium decays of $N_1$ at temperatures $T\ll M_1$. Moreover, taking $M_1\gg 10^{12}\,{\rm GeV}$, charged-lepton flavor effects are unimportant (see, e.g.,~Ref.~\cite{Blanchet:2012bk,leptogenesis:A01} and \sref{Ch4:sec4.2}) and the RH neutrinos are non-relativistic during the production of the asymmetry.

In order to estimate this asymmetry, we need to describe the evolution of the net lepton number $n_{\Delta \ell}$ (only the left-handed SM leptons are accounted for) and the number density of RH neutrinos $n_{N_1}$. We start with the coupled system of \emph{semi-classical} Boltzmann equations for the phase-space distribution functions of the various species $f_{a}(t,\mathbf{X}_a,\mathbf{p}_a)$, whose single-particle energies are \smash{$\omega_{a}(\mathbf{p}_a)=\sqrt{\mathbf{p}_a^2+m_{a}^2}$}. In a Friedmann-Lema\^{i}tre-Robertson-Walker Universe with Hubble rate $H$, there is no dependence on the spatial coordinate $\mathbf{X}_a$, and the semi-classical Boltzmann equations have the form~(see, e.g., Refs.~\cite{Davidson:2008bu} and~\cite{Luty:1992un}):
\begin{align}
&\omega_{a}\frac{\partial f_{a}}{\partial t}\:-\:H\mathbf{p}_{a}^2\frac{\partial f_{a}}{\partial \omega_{a}}\ =\ -\:\frac{1}{2}\sum_{a\mathcal{X}\leftrightarrow \mathcal{Y}}\int{\rm d}\Pi_{\mathcal{X}}\,{\rm d}\Pi_{\mathcal{Y}}\;(2\pi)^4\delta^{4}(p_{a}+p_{\mathcal{X}}-p_{\mathcal{Y}})\nonumber\\&\qquad \times\:\Big[f_{a}f_{\mathcal{X}}|\mathcal{M}(a\mathcal{X}\to \mathcal{Y})|^2F_{\mathcal{Y}}-f_{\mathcal{Y}}|\mathcal{M}(\mathcal{Y}\to a \mathcal{X})|^2F_aF_{\mathcal{X}}\Big]\;.
\end{align}
Our notation is as follows: \smash{$f_{\!\mathcal{A}}(t,\{\mathbf{p}\}) \equiv {\scriptstyle \prod}_{i\in\mathcal{A}}\,f_{i}(t,\mathbf{p}_i)$} is the distribution function of the multi-particle \emph{initial} state \smash{$\mathcal{A}=\mathcal{X},\mathcal{Y}$}; \smash{$F_{\!\mathcal{B}}(t,\{\mathbf{p}\}) \equiv {\scriptstyle \prod}_{i\in\mathcal{B}}\,(1\pm f_{i}(t,\mathbf{p}_i))$} contains the Bose-enhancement ($+$) or Pauli-blocking factors ($-$), due to the quantum statistics of the \emph{final} states $\mathcal{B}=a,\mathcal{X},\mathcal{Y}$; \smash{$\mathcal{M}(a\mathcal{X}\to \mathcal{Y})$} is the matrix element for the process \smash{$a\mathcal{X}\to\mathcal{Y}$}; \smash{$p_{\mathcal{A}}^{\mu} = {\scriptstyle \sum}_{i\in\mathcal{A}}\,p_{i}^{\mu}$}; and \smash{${\rm d}\Pi_{\mathcal{A}} \equiv \sigma_{\!\mathcal{A}}\,{\scriptstyle \prod}_{i\in\mathcal{A}}\,\frac{{\rm d}^4p_{i}}{(2\pi)^4}\,2\pi\theta(p^0_{i})\delta(p_{i}^2-m_{i}^2)$} is the Lorentz-invariant phase-space measure, in which we include a symmetry factor $\sigma_{\!\mathcal{A}}$.

The semi-classical Boltzmann equations above can be recast in a simpler form by making the assumptions of: (i) kinetic (but not chemical) equilibrium and (ii) Maxwell-Boltzmann statistics (valid for the non-relativistic RH neutrinos and leading to a 10\% error for the relativistic species~\cite{Luty:1992un}). These approximations allow us to assume classical statistics, wherein $F_{\mathcal{B}}\approx 1$, and the distribution functions can be written in the form
\begin{equation}
f_a(t,\mathbf{p}_a)\ =\ \frac{n_a(t)}{n^{\rm eq}_a}\,e^{-\omega_a(\mathbf{p}_a)/T}\;,\qquad n_a(t)\ =\ g_a\int\!\frac{{\rm d}^3\mathbf{p}_a}{(2\pi)^3}\,f_a(t,\mathbf{p}_a)\;,
\end{equation}
where the number density $n_a(t)$ has the equilibrium form
\begin{equation}
n_a^{\rm eq}\ =\ g_a\int\!\frac{{\rm d}^3\mathbf{p}_a}{(2\pi)^3}\,e^{-\omega_a(\mathbf{p}_a)/T}\ =\ \frac{g_az_a^2T^3}{2\pi^2}\,\mathcal{K}_2(z_a)\;.
\end{equation}
The factor $g_a$ counts the number of degenerate internal degrees of freedom, $z_a\equiv m_a/T$ and $\mathcal{K}_n$ is the $n$-th order modified Bessel function of the second kind. We are now able to integrate over the Lorentz-invariant phase space of the species $a$ to obtain the \emph{thermally-averaged} Boltzmann equation
\begin{equation}
\label{basics_rateeq}
\dot {n}_a(t)\:+\:3Hn_a(t)\ =\ -\:\sum_{a\mathcal{X}\leftrightarrow \mathcal{Y}}\bigg[\frac{n_a n_{\mathcal{X}}}{n_a^{\rm eq}n^{\rm eq}_{\mathcal{X}}}\,\gamma(a\mathcal{X}\to \mathcal{Y})\:-\:\frac{n_{\mathcal{Y}}}{n_{\mathcal{Y}}^{\rm eq}}\,\gamma(\mathcal{Y}\to a\mathcal{X})\bigg]\;,
\end{equation}
where the $\gamma(a\mathcal{X}\to \mathcal{Y})$ are the thermally-averaged rates:
\begin{equation}
\gamma(a\mathcal{X}\to \mathcal{Y})\ =\ \int{\rm d}\Pi_a\,{\rm d}\Pi_{\mathcal{X}}\,{\rm d}\Pi_{\mathcal{Y}}\;(2\pi)^4\delta^{4}(p_a+p_{\mathcal{X}}-p_{\mathcal{Y}})f_a^{\rm eq}f_{\mathcal{X}}^{\rm eq}|\mathcal{M}(a\mathcal{X}\to \mathcal{Y})|^2\;.
\end{equation}
The relevant processes for the $N_1$-dominated scenario of thermal leptogenesis are the decays and inverse decays of the lightest RH neutrino, as well as the lepton-number-violating scattering processes that it mediates.

In the radiation era, the cosmic time $t$ can be related to the variable $z\equiv M_1/T$ via $t=z^2/(2H_N)$, where $H_N\equiv H(z=1)$. We can then recast the rate equations in still simpler form by introducing the yields 
\begin{equation}
Y_{N_1}\ =\ \frac{n_{N_1}}{s}\qquad \text{and}\qquad Y_{\Delta \ell}\ =\ \sum_{\alpha}Y_{\Delta \ell_{\alpha}}\ =\ \sum_{\alpha}\frac{n_{\ell_{\alpha}}-\bar{n}_{\ell_{\alpha}}}{s}\;,
\end{equation}
where $s=2\pi^2g_{*}T^3/45$ is the entropy density of the $g_{*}$ effective degrees of freedom. For the heavy-neutrino yield, the evolution is dominated by decays and inverse decays, and we obtain the rate equation
\begin{equation}
\frac{s H_N}{z}\,\frac{{\rm d}Y_{N_1}}{{\rm d}z}\ =\ -\:\bigg(\frac{Y_{N_1}}{Y_{N_1}^{\rm eq}}-1\bigg)\sum_{\alpha}\gamma^{N_1}_{\ell_{\alpha} \phi}\;,
\end{equation}
where we have defined the CP-even rates $\gamma^{\mathcal{X}}_{\mathcal{Y}}\ \equiv\ \gamma(\mathcal{X}\to\mathcal{Y})\:+\:\gamma(\bar{\mathcal{X}}\to \bar{\mathcal{Y}})$. For the total lepton asymmetry, we must consider both the heavy-neutrino decays and inverse decays, and the contributions of $s$-channel $\Delta L=2$ scatterings. The resulting rate equation can be written in the form\footnote{The rate equation can be recast in terms of $B-L$, the difference between the baryon and lepton number, which is conserved by the spectator processes. Here, we instead emphasize that we have dealt explicitly with only the doublet leptons and right-handed neutrinos. This notation is used later in~\sref{Ch4:sec4.2}.}
\begin{align}
\label{basics_YDeltaL1}
\frac{s H_N}{z}\,\frac{{\rm d}Y_{\Delta \ell}}{{\rm d}z}\ &=\ \bigg(1+\frac{Y_{N_1}}{Y_{N_1}^{\rm eq}}\bigg)\sum_{\alpha}\delta \gamma^{N_1}_{\ell_{\alpha}\phi}\:-\:\frac{1}{2}\,\sum_{\alpha}\frac{Y_{\Delta \ell_{\alpha}}}{Y_{\ell}^{\rm eq}}\,\gamma^{N_1}_{\ell_{\alpha}\phi}\nonumber\\&\qquad -\:2\sum_{\alpha,\beta}\delta \gamma^{\prime\ell_{\alpha}\phi}_{\phantom{\prime}\bar{\ell}_{\beta}\bar{\phi}}\:-\:\sum_{\alpha,\beta}\frac{Y_{\Delta \ell_{\alpha}}}{Y_{\ell}^{\rm eq}}\,\gamma^{\prime\ell_{\alpha}\phi}_{\phantom{\prime}\bar{\ell}_{\beta}\bar{\phi}}\;,
\end{align}
where we have introduced the CP-odd rates $\delta \gamma^{\mathcal{X}}_{\mathcal{Y}}\ \equiv\ \gamma(\mathcal{X}\to\mathcal{Y})\:-\:\gamma(\bar{\mathcal{X}}\to \bar{\mathcal{Y}})$.

Looking more closely at the term proportional to the CP-odd decay rate $\delta \gamma^{N_1}_{\ell_{\alpha}\phi}$, it would appear that the asymmetry does not vanish, as it should, when the RH neutrinos are in equilibrium. It is for this reason that the $s$-channel $\Delta L = 2$ scattering terms appear with a prime: we must only include those scattering terms that do not count processes already accounted for through the decay and inverse decay terms. This process is referred to as Real Intermediate State (RIS) subtraction~\cite{Kolb:1979qa}. 

In order to understand this procedure, it is convenient to consider the process $\ell_{\alpha}\phi\to\tilde{N}\to\bar{\ell}_{\beta}\bar{\phi}$ mediated by a heavy sneutrino $\tilde{N}$ (rather than a RH neutrino $N$)~\cite{Pilaftsis:2003gt}. The squared matrix element for this process has the form
\begin{equation}
\label{basics_lphitolphi}
|\mathcal{M}(\ell_{\alpha}\phi\to\bar{\ell}_{\beta}\bar{\phi})|^2\ =\ \frac{|\mathcal{M}(\ell_{\alpha}\phi\to\tilde{N})|^2\,|\mathcal{M}(\tilde{N}\to\bar{\ell}_{\beta}\bar{\phi})|^2}{(p^2-M^2)^2+(M\Gamma)^2}\;.
\end{equation}
Making a pole-dominance approximation~\cite{Pilaftsis:2003gt} for the squared modulus of the sneutrino propagator, the RIS contribution can be written
\begin{equation}
\label{basics_lphitolphiRIS}
|\mathcal{M}_{\rm RIS}(\ell_{\alpha}\phi\to\bar{\ell}_{\beta}\bar{\phi})|^2\ =\ \frac{\pi}{M\Gamma}\,\theta(\sqrt{p^2})\,\delta(p^2-M^2)|\mathcal{M}(\ell_{\alpha}\phi\to\tilde{N})|^2\,|\mathcal{M}(\tilde{N}\to\bar{\ell}_{\beta}\bar{\phi})|^2\;,
\end{equation}
from which $\gamma_{\rm RIS}(\ell_{\alpha}\phi\to\bar{\ell}_{\beta}\bar{\phi})=\gamma(\ell_{\alpha}\phi\to\bar{\ell}_{\beta}\bar{\phi})-\gamma'(\ell_{\alpha}\phi\to\bar{\ell}_{\beta}\bar{\phi})$ can be obtained on thermal averaging. Although technically more involved, a completely equivalent result can be obtained for the exchange of a RH neutrino (see Ref.~\cite{Pilaftsis:2003gt}). Putting everything together, it can then be shown (assuming only $N_1$ exchange) that~\cite{Kolb:1979qa,Pilaftsis:2003gt}
\begin{equation}
\sum_{\beta}\delta \gamma^{\prime\ell_{\alpha}\phi}_{\phantom{\prime}\bar{\ell}_{\beta}\bar{\phi}}\ =\ \delta \gamma^{N_1}_{\ell_{\alpha}\phi}\:+\:\mathcal{O}(\lambda^4)\;.
\end{equation}
On substituting this into Eq.~\eqref{basics_YDeltaL1}, we see that the scattering terms conspire to change the overall sign of the first term, such that the asymmetry vanishes in the equilibrium limit, as it should. Keeping only the (RIS corrected) decay and inverse decay terms, the equation for the asymmetry now reads
\begin{equation}
\label{basics_YDeltaL2}
\frac{s H_N}{z}\,\frac{{\rm d}Y_{\Delta \ell}}{{\rm d}z}\ =\ -\:\bigg(1-\frac{Y_{N_1}}{Y_{N_1}^{\rm eq}}\bigg)\sum_{\alpha}\delta \gamma^{N_1}_{\ell_{\alpha}\phi}\:-\:\frac{1}{2}\,\sum_{\alpha}\frac{Y_{\Delta \ell_{\alpha}}}{Y_{\ell}^{\rm eq}}\,\gamma^{N_1}_{\ell_{\alpha}\phi}\;.
\end{equation}

The CP-odd rate $\delta \gamma^{N_i}_{\ell_{\alpha}\phi}$ can be written in terms of the CP-even rates $\gamma^{N_i}_{\ell_{\alpha}\phi}$ via \smash{$\delta \gamma^{N_i}_{\ell_{\alpha}\phi}\ =\ \epsilon_{i\alpha}{\scriptstyle \sum}_{\beta}\gamma^{N_i}_{\ell_{\beta}\phi}$},
where $\epsilon_{i\alpha}$ are the per-flavor CP asymmetries
\begin{equation}
\epsilon_{i\alpha}\ =\ \frac{\Gamma(N_i\to\ell_{\alpha}\phi)-\Gamma(N_i\to \bar{\ell}_{\alpha}\bar{\phi})}{\sum_{\beta}\big[\Gamma(N_i\to\ell_{\beta}\phi)+\Gamma(N_i\to \bar{\ell}_{\beta}\bar{\phi})\big]}\;.
\label{CPdef1}
\end{equation}
For the $N_1$-dominated scenario and in the one-flavour approximation, the rate equations become (see, e.g., Refs.~\cite{Buchmuller:2004nz,Fong:2013wr})
\begin{gather}
\frac{{\rm d}Y_{N_1}}{{\rm d}z}\ =\ -\:D_1\big(Y_{N_1}-Y_{N_1}^{\rm eq}\big)\;,\\
\frac{{\rm d}Y_{\Delta \ell}}{{\rm d}z}\ =\ \epsilon_1 D_1\big(Y_{N_1}-Y_{N_1}^{\rm eq}\big)\:-\:WY_{\Delta \ell}\;,
\end{gather}
where $\epsilon_1\equiv \sum_{\alpha}\epsilon_{1\alpha}$. We can now identify the decay terms, proportional to
\begin{equation}
\label{basics_D}
D_1\ \equiv\ \frac{z}{H_N\,n_{N_1}^{\rm eq}}\,\sum_{\alpha}\gamma^{N_1}_{\ell_{\alpha}\phi}\ =\ K_1\,z\,\frac{\mathcal{K}_1(z)}{\mathcal{K}_2(z)}\;,
\end{equation}
and the washout due to inverse decays, proportional to
\begin{equation}
W\ \equiv\ \frac{1}{2}\,D_1\,\frac{Y_{N_1}^{\rm eq}}{Y_{\ell}^{\rm eq}}\;.
\end{equation}

The factor $K_1\equiv\ \Gamma_1/H_N$, appearing in Eq.~\eqref{basics_D}, is called the decay parameter. For the present scenario, $K_1\gg 1$, and we are in the so-called strong washout regime. In this case, $Y_{N_1}\approx Y_{N_1}^{\rm eq}$, and any initial asymmetry present before or generated during the production phase of the lightest RH neutrinos (at temperatures $T> M_1$) is completely washed out. We can then estimate the final asymmetry by setting the source and washout terms equal to one another~\cite{Fong:2013wr, Fong:2010up}:
\begin{equation}
Y_{\Delta \ell}\ \approx\ -\:\frac{\epsilon_1}{W}\,\frac{{\rm d}Y_{N_1}}{{\rm d}z}\ \approx\ -\:\frac{\epsilon_1}{W}\,\frac{{\rm d}Y_{N_1}^{\rm eq}}{{\rm d}z}\ =\ \frac{2}{z\,K_1}\,\epsilon_1 \,Y_{\ell}^{\rm eq}\;,
\end{equation}
where $Y_{\ell}^{\rm eq}\ =\ 15/(4\pi^2g_*)$. We see that the final asymmetry is proportional to the CP-violating parameter and inversely proportional to the decay factor $K$; if the CP-violating parameter is too small or the washout too strong, the final asymmetry is suppressed.

\begin{figure}[t!] \centering
\includegraphics[scale=0.55]{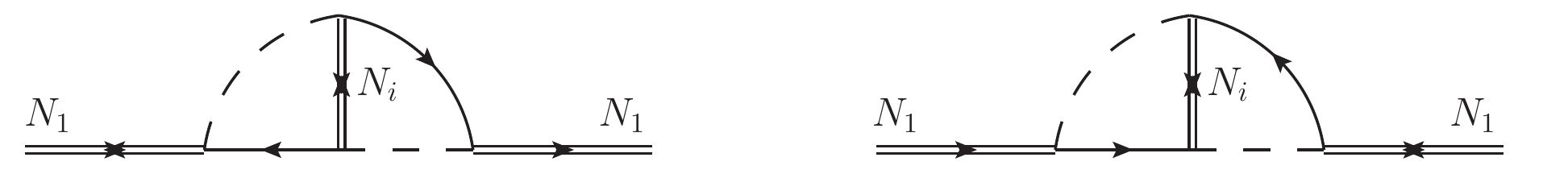}
\caption{Self-energy diagrams for the lightest Majorana neutrino, $N_{1}$: direct contributions.
Solid double lines stand for RH neutrinos, solid lines for SM lepton doublets and dashed lines for Higgs bosons. 
The neutrino propagator with forward arrow corresponds to $\langle 0| T(N_k \bar{N}_k) |0\rangle$, 
whereas the neutrino propagators with forward-backward arrows correspond to 
$\langle 0| T(N_k N_k) |0\rangle$ or $\langle 0| T(\bar{N}_k \bar{N}_k)|0 \rangle$.} 
\label{fig:self1} 
\end{figure}

\begin{figure}[t!]  \centering
\includegraphics[scale=0.49]{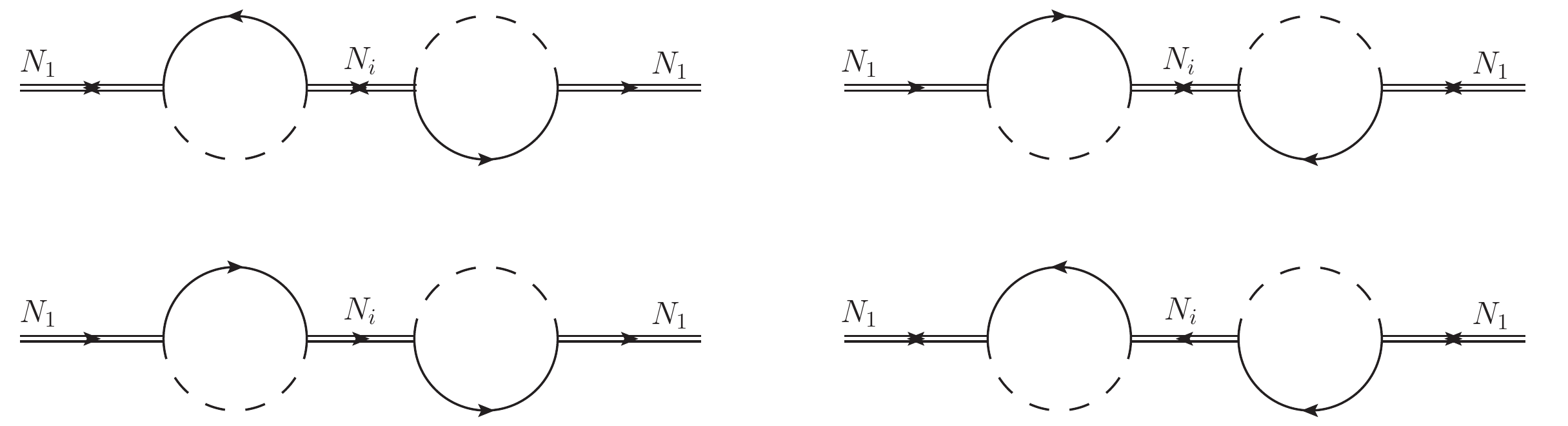}
\caption{Self-energy diagrams for the lightest Majorana neutrino, $N_{1}$: indirect contributions.}
\label{fig:self2} 
\end{figure}

At tree level, we have $\Gamma(N_i\to \ell_{\alpha}\phi)=\Gamma(N_i\to \bar{\ell}_{\alpha}\bar{\phi})$, and the CP-violating parameter is zero. The sources of CP violation instead arise through the interference of the tree-level and one-loop processes, pictured in \fref{fig:self1} and \fref{fig:self2}. The per-flavour CP-violating parameters are then given by~\cite{Fukugita:1986hr,Covi:1996wh}
\begin{align}
\label{basics_CPepsilons}
\epsilon_{i\alpha}\ &=\ \frac{1}{8\pi}\,\sum_{j\,\neq\,i}\frac{{\rm Im}\big[\lambda^*_{\alpha i}(\lambda^{\dag}\lambda)_{ij}\lambda_{\alpha j}\big]}{|\lambda_i|^2} \, \xi\bigg(\!0,\frac{M_j^2}{M_i^2}\bigg)\nonumber\\&\qquad+\:\frac{1}{8\pi}\,\sum_{j\,\neq\,i}\frac{{\rm Im}\big\{\lambda_{\alpha i}^*\big[(\lambda^{\dag}\lambda)_{ij}M_j+M_i(\lambda^{\dag}\lambda)_{ji}\big]\lambda_{\alpha j}\big\}}{|\lambda_i|^2}\frac{M_i}{M_i^2-M_j^2}\;,
\end{align}
where
\begin{equation}
\xi(b,x)\ =\ \sqrt{x}\bigg[1+\frac{b}{1-x}\:-\:(1+x)\ln\frac{1+x}{x}\bigg]
\label{explicit_vertex_CP}
\end{equation}
and the mass splittings of the RH neutrinos are assumed to be large compared to their characteristic decay width.

The first term on the right-hand side of Eq.~\eqref{basics_CPepsilons} corresponds to the so-called \emph{direct} or $\varepsilon'$-type CP violation, which arises from the interference between the tree-level diagram and the one-loop vertex correction (see Fig.~\ref{fig:self1}). The terms on the second line of Eq.~\eqref{basics_CPepsilons} correspond to the so-called \emph{indirect} or $\varepsilon$-type CP violation, which arises from the interference between the tree-level diagrams and the one-loop self-energy corrections (see Fig.~\ref{fig:self2}).  For mass splittings that are small compared to the characteristic decay width, the $\varepsilon$-type CP violation is enhanced and can dominate over the $\varepsilon'$-type CP violation, leading to so-called resonant leptogenesis (see, e.g., Ref.~\cite{Pilaftsis:2003gt} and chapters~\cite{leptogenesis:A01} and \cite{leptogenesis:A03} of this review). Conversely, when the mass splittings are comparable to the masses themselves, the $\varepsilon$- and $\varepsilon'$-type CP violation are of similar magnitude. In the one-flavour regime, the lattermost term on the second line of Eq.~\eqref{basics_CPepsilons} (proportional to $M_i^2$) does not contribute to the asymmetry, and we have
\begin{equation}
\epsilon_i\ =\ \sum_{\alpha}\epsilon_{i\alpha}\ =\ \frac{1}{8\pi}\sum_{j\,\neq\, i}\frac{{\rm Im}[(\lambda^{\dag}\lambda)^2_{ij}]}{|\lambda_i|^2}\,\xi\bigg(\!1,\frac{M_j^2}{M_i^2}\bigg)\;.
\end{equation}

The remainder of this chapter will focus on the thermal field theoretic derivations of the key ingredients that we have introduced in this section. In so doing, we will see how the current state-of-the-art goes beyond the simplistic analysis and approximations that we have detailed above.

\section{The right-handed neutrino production rate}\label{Ch4:sec2}
The purpose of this section is to consider in-vacuum and thermal corrections  to the RH neutrino production rate in the simplest realization of thermal leptogenesis, described in \sref{basics_lep}. 
The neutrino production rate can be understood as the space-time average of the rate at
which the thermal plasma creates quanta of the lightest RH neutrino.
In thermal equilibrium, the creation rate equals the destruction rate. Even though we are interested in out-of-equilibrium dynamics to address the lepton asymmetry generation, the RH neutrino production rate can be extracted in terms of an equilibrium distribution function. 
We label the production rate as $\gamma_{N_1} \equiv \gamma$, and we express it as the imaginary part of a retarded self-energy, $\Pi_{R}$, as follows  \cite{Salvio:2011sf, Laine:2011pq}
\begin{equation}
\gamma= 2 \int \frac{d^3 \bm{k}}{(2\pi)^3}  \, \frac{{\rm{Im}} \, \Pi_R(\omega)}{\omega} \,  f_{N}^{{\rm{eq}}}(\omega) \, ,
\label{prod_right}
\end{equation}
where $\omega=\sqrt{k^2+M^2}$ with $k=|\bm{k}|$,  $f_{N}^{{\rm{eq}}}$ is the Fermi-Dirac distribution, the factor of two is due to the spin polarization. Moreover a neutrino width can be defined as
\begin{equation}
\Gamma(\omega) \equiv \frac{{\rm{Im}}\Pi_R(\omega)}{\omega} \, ,
\end{equation} 
that, at leading order in the couplings, at zero temperature and in the neutrino rest frame, simply reads  $\Gamma^{T=0}_{\hbox{\tiny LO}}=M |\lambda|^2/(8 \pi)$.
The neutrino width is a more fundamental object than the neutrino production rate, see \eref{prod_right}, and we are going to review its radiative and thermal corrections in what follows.
\subsection{In-vacuum and thermal corrections in the non-relativistic regime}\label{Ch4:sec2.1}
In the simplest realization of thermal leptogenesis, $M$ is much larger than the electroweak scale and then a great simplification can be obtained by taking all SM masses to be negligible with respect to the heavy-neutrino mass.\footnote{At $T\gsim 160$ GeV the Higgs mechanism is not operative, hence SM particles can only get thermal masses of order $gT$. Nevertheless, the latter would be much smaller than the RH neutrino mass scale, $M \gg T \gg gT$.} However, some processes contributing to $\Gamma$ are  infrared divergent in this limit. For example,  the contribution coming from
$2\rightarrow 2$ scatterings (such as $AN \to \ell \phi$,
where $A$ is any SM vector) are not infrared finite. The corresponding expressions 
are lengthy because some sort of regulator, e.g. thermal masses, has to be introduced for the infrared divergences\cite{Barbieri:1999ma,Giudice:2003jh}. Only summing all contributions at a given order in perturbation theory leads to a finite result\cite{Salvio:2011sf}. For example, the process  $AN \to \ell \phi$ has to be considered together with 3-body decays, such as $N\to \ell \phi A$, and virtual corrections in order to obtain a finite result.

The relevant SM couplings involved in the next-to-leading order (NLO) calculation of $\Gamma$ are the SU(2)$_L$ and U(1)$_Y$ gauge couplings $g$ and $g'$ respectively, the top Yukawa  coupling $\lambda_t=\sqrt{2}m_t/v$
and the Higgs self interaction $\lambda_h=m_h^2/(2v^2)$, where $m_h$ is the zero temperature Higgs mass. The NLO  expression for $\Gamma$ in the rest frame of $N$ is \cite{Salvio:2011sf, Laine:2011pq} 
\begin{equation}
 \Gamma = \label{eq:final}
 \frac{M |\lambda|^2}{8\pi}
 \bigg[1 +\underbrace{ \frac{29}{128\pi^2}(3 g^2 + g'^2)}_{3\% (2.5\%)}
 -\underbrace{ \frac{21\lambda_t^2}{32\pi^2}}_{5\%(2\%)}
-  \lambda_h \frac{T^2}{M^2}\bigg].
\end{equation} 
Only the leading thermal correction is displayed at this stage.  Also, the numerical values for couplings defined in the $\overline{\text{MS}}$ scheme and  renormalized at the weak scale (at $10^{10}$ GeV) are shown, having 
fixed $\lambda_h=m_h^2/(2v^2)$ and assumed $m_h = 125$ GeV. 

The first three terms in \eref{eq:final} are the zero-temperature part of the width and radiative corrections are included that correspond to virtual corrections to $1\to 2$ decays of $N$ as well as real corrections
(i.e.~$2\to 2$ scatterings, $1\to 3$ decays). They can be obtained with zero-temperature QFT methods.
The temperature correction in \eref{eq:final}  can be obtained by using the real-time formalism of thermal field theory. In the limit where $\lambda$ is small with respect to the relevant SM couplings, one can compute $\Gamma$  from the imaginary part   of the $N$-propagator in the thermal plasma, computed by explicitly summing all possible ``cuts" of the relevant Feynman diagrams. The finite-temperature cuts can be obtained by the so-called Kobes-Semenoff rules, which yield the absorptive parts of Feynman diagrams\cite{LeBellac}, and generalize the cutting rules valid at $T = 0$.
\begin{figure}[t!]
\centering
\includegraphics[width=9.5cm,clip]{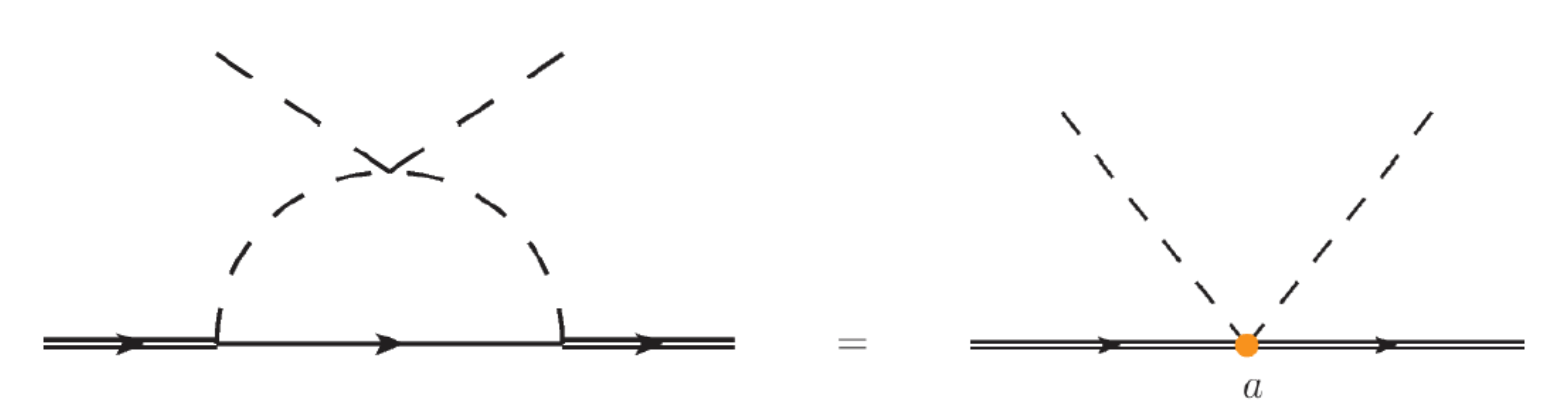}
\caption{One-loop matching condition for the neutrino-Higgs coupling. Double lines are RH neutrino propagators, single lines lepton propagators and dashed lines Higgs propagators.}
\label{fig-matching}   
\end{figure}

The neutrino width has recently been recast in terms of a non-relativistic EFT for heavy Majorana fermions that may be dubbed Heavy Majorana Effective Theory (HMET) 
in analogy with the Heavy Quark Effective Theory\cite{Biondini:2013xua}. Indeed one may implement, at the Lagrangian level, the separation of the energy scales, $M \gg T$. The advantages of an EFT treatment for heavy particles can be summarized as follows. Firstly the EFT makes manifest, already at the 
Lagrangian level, the non-relativistic nature of the Majorana particle.
Secondly it allows to separate the computation of radiative and thermal corrections: 
radiative corrections are computed setting $T=0$ and contribute to 
the Wilson coefficients of the EFT, whereas thermal corrections are computed in the EFT as small 
corrections affecting the propagation of the non-relativistic RH neutrinos in the thermal medium.
Finally, the power counting of the EFT allows a rather 
transparent organization of the calculation, leading to several simplifications 
that would not be obvious at the level of the relativistic thermal field theory.

At  an energy scale  much smaller  than $M$, the  low-energy modes of the  Majorana neutrino can be described by a field
$\psi$, whose effective interactions with the SM particles are encoded in the EFT \cite{Biondini:2013xua}
\begin{equation}
\mathcal{L}_{\hbox{\tiny EFT}}= \mathcal{L}_{\hbox{\tiny SM}} + \mathcal{L}_{\hbox{\tiny $\psi$}}= \mathcal{L}_{\hbox{\tiny SM}} + \bar{\psi} \left(iv \cdot \partial + \frac{i\Gamma^{T=0}}{2} \right) \psi + \frac{\mathcal{L}^{(1)}}{M}
+\frac{\mathcal{L}^{(2)}}{M^2}
+\frac{\mathcal{L}^{(3)}}{M^3}
+\dots \,.
\label{EFT1_maj}
\end{equation}
The power counting of the EFT indicates that the leading operators responsible 
for the neutrino thermal decay are dimension-five operators contributing to $\mathcal{L}^{(1)}$.
The symmetries of the EFT allow for only one possible dimension-five operator, which reads 
\begin{equation}
\mathcal{L}^{(1)}= a \; \bar{\psi} \psi \, \phi^{\dagger} \phi\,. 
\label{EFTL1}
\end{equation}
This operator describes the scattering of Majorana neutrinos with Higgs particles. By dimensional arguments, its contribution to a thermal width has to scale as $T^2/M$.
The Wilson coefficient $a$ is fixed at one loop by the matching condition 
shown in \fref{fig-matching}. The left-hand side stands for an (in-vacuum) 
diagram in the fundamental theory in \eref{Lag1_full}, whereas the right-hand side for 
an (in-vacuum) diagram in the EFT in \eref{EFT1_maj}. 
For the decay width, only the imaginary part of the matching coefficient is relevant; one finds  ${\rm{Im}}\,a  = -3 |\lambda|^{2}\lambda_h /(8\pi) $.

\begin{figure}[t!]
\centering
\includegraphics[width=5.5 cm,clip]{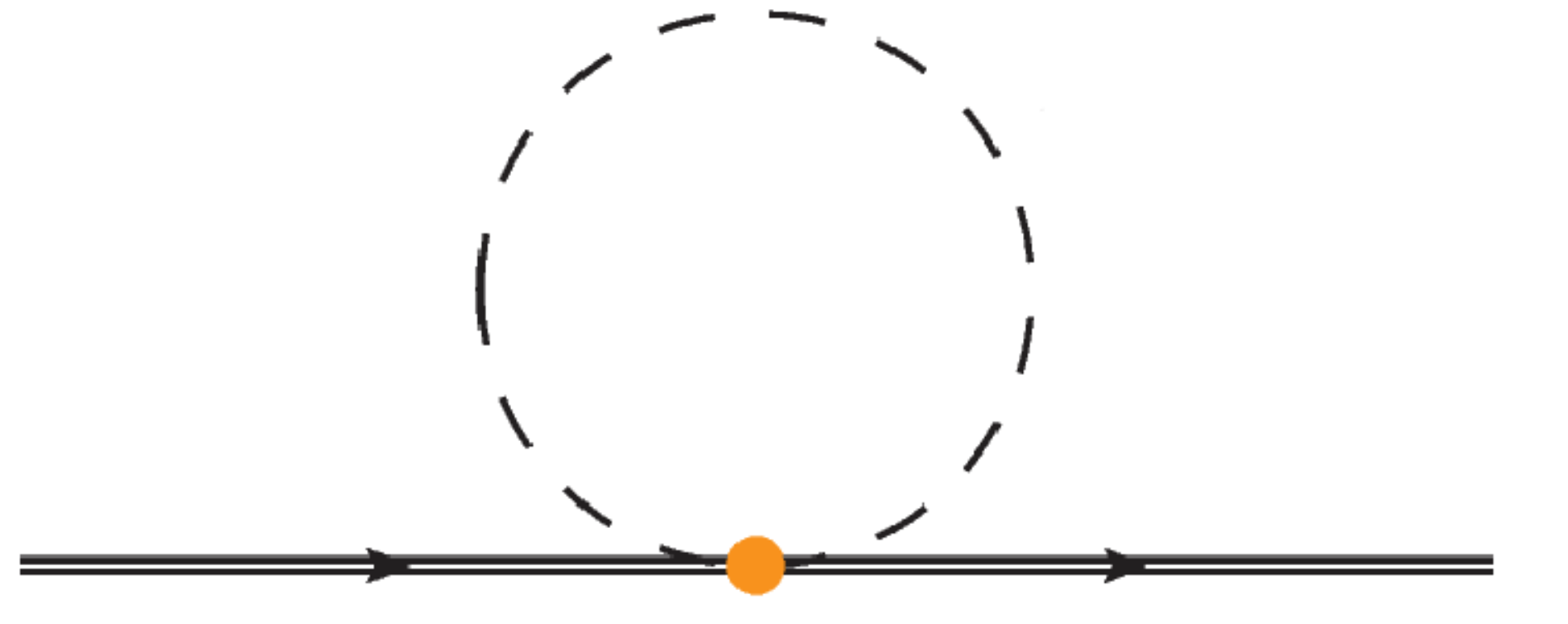}
\caption{Higgs tadpole contribution to the neutrino thermal width.}
\label{fig-loop}   
\end{figure}

The thermal width induced by \eref{EFTL1} can be computed from the tadpole  diagram 
shown in \fref{fig-loop}, where the dashed line 
has to be understood now as a thermal Higgs propagator.
The leading thermal width reads~\cite{Salvio:2011sf,Laine:2011pq,Biondini:2013xua},
\begin{equation}
\Gamma_\phi^T = 2 \frac{{{\rm Im} \, a}}{M} \langle \phi^{\dagger}(0) \phi(0) \rangle_{T}  
=  -\frac{|\lambda|^{2}M}{8\pi} \lambda_h \left( \frac{T}{M} \right)^{2}\,,
\label{leadingwidth}
\end{equation}
where $ \langle \phi^{\dagger}(0) \phi(0) \rangle_{T}$ stands for the 
thermal condensate of the field $\phi$. The above expression of $\Gamma_\phi^T$ coincides with the thermal part of \eref{eq:final}.
Note that in the EFT the calculation has split into a one-loop matching, shown in 
\fref{fig-matching}, which can be done in vacuum, and the calculation 
of a one-loop tadpole diagram, shown in \fref{fig-loop}, which is done in 
thermal field theory. 
In a similar fashion, one can calculate $T/M$-suppressed corrections to the thermal decay width.
Only dimension-seven operators contribute to the width at next order in  $T/M$, and they are included in $\mathcal{L}^{(3)}$. They describe couplings of the 
Majorana neutrino to Higgs bosons, leptons, quarks and gauge bosons respectively.
Finally, the thermal width at first order in the SM couplings and at order $(T/M)^4$ (and in the neutrino rest frame) 
reads~\cite{Laine:2011pq, Biondini:2013xua}
\begin{eqnarray}
&& \Gamma^T 
=  \frac{|\lambda|^{2}M}{8\pi}\left[-\lambda_h \, \left( \frac{T}{M} \right)^{2} 
- \frac{\pi^{2}}{80}\,(3g^{2}+g'^{\,2}) \,\left( \frac{T}{M} \right)^{4}
- \frac{7\pi^{2}}{60}\,|\lambda_{t} |^{2}\,\left( \frac{T}{M} \right)^{4} \right]\,.
\label{gamma_nonrel}
\end{eqnarray}
Corrections due to the RH neutrino motion are included to the above accuracy in Ref.\cite{Biondini:2013xua} within the EFT framework, whereas they are fully taken into account in Ref.\cite{Laine:2011pq}, together with zero-temperature quantum corrections at leading order in the SM couplings. 
\subsection{Right-handed neutrino production rate in the relativistic regime}\label{Ch4:sec2.2}
After summing over helicities (denoted below by $s = \pm$)
and including lepton chemical potentials (denoted by $\mu_\alpha$), 
the production rate of the $j$-th RH neutrino
with momentum $k$ can be written as (a sum over active lepton flavor is understood)
\begin{eqnarray}
 \sum_s \dot{f}_j(k) & = &  
 |\lambda^{ }_{\alpha j}|^2
 \Bigl\{ 
  \frac{\Gamma^+_{\alpha j}}{e^{(\omega_j - \mu_\alpha)/T} + 1}  
 +   
  \frac{\Gamma^-_{\alpha j}}{e^{(\omega_j + \mu_\alpha)/T} + 1}  
 \Bigr\} 
 + O(\lambda^4)
 \;, \label{prod} \\ 
  \Gamma_{\alpha j}^{\pm}
  & \equiv &  \frac{1}{2 \omega_j} 
  \tr \Bigl[ \bsl{K}_{\! j}\,
  \rho_\alpha (\pm K_j) \Bigr] 
 \;. 
\end{eqnarray}
Here $K_j = (\omega_j,\bm{k})$, $\omega_j \equiv \sqrt{k^2 + M_j^2}$, and 
\begin{equation}
 \rho_\alpha (K_j) \equiv 
 \int_{-\infty}^{\infty} \! {\rm d}t \int_{\bm{x}} 
 e^{i (\omega_j t - \bm{k}\cdot\bm{x})}
 \Bigl\langle  \Bigl\{ 
  (\phi^{c\dagger} \ell_\alpha)(t,\bm{x}) , (\bar{\ell}_\alpha \phi^c)(0) 
 \Bigr\} \Bigr\rangle 
 \label{rho_laine}
\end{equation}
is the so-called spectral function 
(twice the imaginary part of a retarded correlator\footnote{There exist different conventions for the spectral function, see, e.g.\cite{Laine:2016hma}, where a factor 1/2 appears in \eref{rho_laine}.}). 
The two contributions in \eref{prod} 
originate from leptons and antileptons, 
respectively. If the lepton chemical potentials vanish, 
i.e.\ $\mu_\alpha = 0$, then 
$
  \Gamma_{\alpha j}^{+} =  \Gamma_{\alpha j}^{-}
$.
The production rate is a physically relevant concept as long as the 
distribution function $f_j$ remains much below its equilibrium value,
given by the Fermi-Dirac distribution. 
However, the same coefficients $ \Gamma_{\alpha j}^{\pm} $ also 
govern the behavior of $f_j$ close to 
equilibrium~\cite{Bodeker:2015exa}, 
even though the form of the differential \eref{prod} 
is then different. 

The task now is to compute $ \Gamma_{\alpha j}^{\pm} $ within a 
SM plasma at a temperature $T$. The system has at least 
three different scales that affect the computation. 
One is the SM crossover temperature
$T_c\sim 160$~GeV~\cite{Laine:2015kra,D'Onofrio:2015mpa}, below
which the Higgs mechanism is operative. The other two are 
the temperature $T$ and the Majorana mass $M_j$. It is important
to note that even if we were only interested in leptogenesis, 
then the Higgs phase still needs to be considered, given that
sphaleron processes, which partially convert lepton asymmetries into
baryon asymmetries, only switch off at 
$T \sim 130$~GeV~\cite{D'Onofrio:2014kta}. 
 
The techniques used and the precision of the computations
that has been reached to date depend on the parametric regime considered,
i.e.\ the relations of the three scales mentioned above. 
In the present section, we focus on the relativistic and ultrarelativistic
regimes. The relativistic regime corresponds to $M_j \sim \pi T$, 
and the ultrarelativistic 
to $M_j \lsim gT$, where $g$ denotes the weak gauge coupling. 
For $M_j \sim $ GeV the ultrarelativistic regime 
covers temperatures  $T \gsim 5$~GeV, i.e.\ all temperatures
of interest for leptogenesis. On the other hand, 
for $M_j \gsim $ TeV, both the relativistic and ultrarelativistic
regimes are relevant at the early stages of the evolution, whereas
the very late stages may be addressed with non-relativistic
methods (see \sref{Ch4:sec2.1}). 

Starting with the relativistic regime, the main computational challenge 
is posed by the fact that when two scales are of similar magnitudes, 
$M_j \sim \pi T$, then no kinematic simplifications are possible, 
and both scales need to be retained. At NLO, this implies that computations become technically 
complicated~\cite{Garbrecht:2013gd}. Nevertheless all phase-space
integrals appearing can be reduced into 2-dimensional 
ones~\cite{Laine:2013vpa}, which can subsequently be evaluated
numerically. The full NLO computation for this regime, assuming
$T \gg 160$~GeV, is described in Ref.~\cite{Laine:2013lka}, and it generalizes the non-relativistic result discussed in \sref{Ch4:sec2.1} to a broader temperature range. 


If the temperature is increased so that $\pi T \gg M_j$, the 
relativistic results themselves break down. In this ultrarelativistic
regime, infinite ``resummations'' (two nested resummations, 
Hard Thermal Loop resummation and Landau-Pomeranchuk-Migdal resummation)
are needed in order to obtain even the correct leading-order result. 
For $T > 160$~GeV,
the techniques and results relevant for the ultrarelativistic regime
were worked out in Refs.~\cite{Besak:2010fb,Anisimov:2010gy,Besak:2012qm}.  
They have been extended to $T \lsim 160$~GeV in 
Ref.~\cite{Ghiglieri:2016xye}. Also, a consistent 
way to interpolate between the ultrarelativistic and relativistic 
regimes has been worked out for $T > 160$~GeV~\cite{Ghisoiu:2014ena}, 
so that, for high temperatures, results applicable to any $M_j$ and $\pi T$
are available. Up-to-date numerical results for the rates in the various
parametric regimes can be found through the web page
\url{http://www.laine.itp.unibe.ch/production-midT/}.

One price to pay for the resummations of the ultrarelativistic 
regime is that only the leading order in SM couplings
has been reached so far. 
However, in principle, the NLO level can also be 
attacked; corrections could indeed be large, because they are
only suppressed by $O(g)$ in this regime. In the context of 
particle production rates in a QCD plasma, the
corresponding techniques have been worked out 
in Refs.~\cite{Ghiglieri:2013gia,Ghiglieri:2014kma}. 

The above computations referred to a system 
in which chemical potentials had been set to zero. 
At $T\gsim 160$~GeV, a non-zero lepton 
chemical potential was included in 
Refs.~\cite{Hernandez:2016kel,Ghiglieri:2017gjz}.  
In addition, at low temperatures $T \lsim 1$~GeV, 
chemical potentials play a decisive role~\cite{Shi:1998km}, 
and have been included in 
Refs.~\cite{Ghiglieri:2015jua,Venumadhav:2015pla}. 

Recently, it has been realized that the two helicity states
of a massive RH neutrino, denoted by $s$ and 
summed together in \eref{prod}, 
actually behave quite differently. This may lead to important
physical effects. Rates for the specific helicity states 
were considered at $T < 130$~GeV in Ref.~\cite{Eijima:2017anv} and at
$T > 130$~GeV in Ref.~\cite{Ghiglieri:2017gjz}. A partial earlier 
investigation at low temperatures can be found 
in Ref.~\cite{Lello:2016rvl}, albeit for a toy model 
with a Dirac rather than a Majorana RH neutrino. 

Apart from particle production, the same 
coefficients $\Gamma^\pm_{\alpha j}$ also play a role
for lepton number washout rates~\cite{Bodeker:2014hqa}, which represent
an important ingredient in any leptogenesis computation. In this
case, the coefficients $\Gamma^\pm_{\alpha j}$ come in combination
with so-called lepton number susceptibilities, which have been 
computed up to NLO~\cite{Bodeker:2014hqa} 
and NNLO~\cite{Bodeker:2015zda} at $T > 160$~GeV. 
At $T \sim 160$~GeV the susceptibilities suffer from infrared
divergences related to light Higgs modes and would require a
non-perturbative determination. Leading-order numerical results
for the susceptibilities at $T < 130$~GeV and $T > 130$~GeV can
be found on the web page
\url{http://www.laine.itp.unibe.ch/production-midT/}. A recent discussion about the influence of susceptibilities at $T \sim 130$~GeV can be found in Ref.~\cite{Eijima:2017cxr}. 
\section{CP-violating parameter from right-handed neutrino decays}\label{Ch4:sec3}
As discussed in \sref{basics_lep}, CP-violating parameters are a crucial ingredient entering the rate equations for the lepton-number
asymmetry.
They are related to the CP asymmetries generated by the Majorana neutrinos when decaying into leptons and antileptons carrying a flavor $\alpha$. 
We discuss thermal corrections to the CP asymmetry in the non-relativistic regime and then its connection with the lepton asymmetry.
\subsection{Towards NLO corrections to the CP-violating parameter in the non-relativistic regime}\label{Ch4:sec3.1}
At zero temperature and at zeroth order in the SM couplings, the RH neutrino CP asymmetries have been known for some time\cite{Covi:1996wh,Fong:2013wr}.
They may be computed from the diagrams shown in \fref{fig:self1} and \fref{fig:self2} by cutting through lepton and Higgs-boson lines. 

Contributions from the diagrams of \fref{fig:self1} are sometimes called direct contributions, because they are not resonant in the limit of nearly degenerate neutrinos. 
For the lightest Majorana neutrino, conventionally taken to be of type 1, they give (from \eref{basics_CPepsilons} and \eref{explicit_vertex_CP})  
\begin{equation}
  \epsilon_{1\alpha}^{T=0,{\rm{direct}}} =
  \frac{M_i}{M_1} \left[ 1-\left( 1+\frac{M^2_i}{M^2_1} \right)  
\ln \left( 1+ \frac{M_1^2}{M_i^2} \right) \right] \frac{ {\rm{Im}}\left[ (\lambda^{*}_1\lambda_i)( \lambda^*_{\alpha 1}\lambda_{\alpha i})  \right] }{8 \pi|\lambda_1|^2} .
\label{CPvert}
\end{equation}
Here and in the rest of this section, the neutrino-type index $i$ is understood as summed over all neutrino species different from $1$ ($i > 1$).

Contributions from the diagrams of \fref{fig:self2} are sometimes called indirect contributions, because they are resonant in the limit of nearly degenerate neutrinos (see Ref.~\cite{leptogenesis:A03} for more details).
For the lightest RH neutrino, they give (again from \eref{basics_CPepsilons})
\begin{equation}
  \epsilon_{1\alpha}^{T=0,{\rm{indirect}}} =
  \frac{M_1 M_i}{M_1^2-M_i^2} \frac{ {\rm{Im}}\left[ (\lambda^{*}_1\lambda_i)( \lambda^*_{\alpha 1}\lambda_{\alpha i})  \right] }{8 \pi |\lambda_1|^2} 
+ \frac{M_1^2}{M_1^2-M_i^2} \frac{ {\rm{Im}}\left[ (\lambda_1\lambda^{*}_i)( \lambda^*_{\alpha 1}\lambda_{\alpha i})  \right] }{8 \pi|\lambda_1|^2} .
\label{CPself}
\end{equation}
The distinction between direct and indirect contributions loses significance far away from the degenerate limit.

In a thermal medium, decay widths and CP asymmetry parameters appearing in the rate equations get thermal corrections.
While the RH neutrino thermal decay width is known at first order in the SM couplings (see \sref{Ch4:sec2.2}), 
this is not yet the case for the thermal corrections to the RH neutrino CP asymmetries generated by decays into leptons and antileptons, 
for which only partial results can be found in the literature\cite{Covi:1997dr,Giudice:2003jh}.
Recently, however, the RH neutrino CP asymmetries have been computed in an expansion in $T/M_k$ and at first order in the SM couplings. 
The derivation is based on the HMET which makes the $1/M_k$ expansion explicit at the Lagrangian level and allows an efficient organization of the calculation.
The results may be useful if implemented in non-relativistic rate equations\cite{Bodeker:2013qaa} for the late-time (low-temperature) evolution
of the lepton-number asymmetry and, once a fully relativistic result valid for all temperatures is available, as a non-trivial constraint in the low-temperature regime. 
The results obtained in the HMET hold for temperatures lower than the lightest Majorana neutrino mass and (in the setting of Ref.~\cite{Biondini:2013xua}) larger than the electroweak crossover scale. Including NLO contributions in the CP parameters, more general final states can appear in the heavy-neutrino decays, so that we define  
\begin{equation}
\epsilon_{k\alpha}=  \frac{  \Gamma(N_{k} \to \ell_{\alpha} + X)-\Gamma(N_{k} \to \bar{\ell}_{\alpha}+ X )  }
{\sum_\beta \, \Gamma(N_{k} \to \ell_{\beta} + X )+\Gamma(N_{k} \to \bar{\ell}_{\beta}+ X)} ,
\label{CPdef1_ext}
\end{equation}
where $\Gamma$ is the width of the specified decay process and $X$ represents any other SM particle not carrying a lepton number.

We first review the hierarchical case, where the lightest neutrino mass $M_1$ is much smaller than the other neutrino masses $M_i$, with $i>1$\cite{Biondini:2016arl}. In this case, the full NLO result for the in-vacuum radiative corrections to the CP-violating rates has been derived  recently in Ref.\cite{Bodeker:2017deo}.
Thermal corrections to the CP asymmetry of the lightest neutrino
were computed in terms of an expansion in the Yukawa couplings, SM couplings, $(M_1/M_i)$ and $(T/M_1)$;
they read
\begin{eqnarray}
\epsilon_{1\alpha}^{T} &=& 
-\frac{3}{16 \pi} \frac{M_1}{M_i} \, \frac{ {\rm{Im}}\left[ (\lambda^{*}_1\lambda_i)( \lambda^*_{\alpha 1}\lambda_{\alpha i})  \right]}{|\lambda_1|^2} 
\left[   \left(  -\frac{5}{3} \lambda_h + \frac{2g^2+g'^2}{12} \right) \left( \frac{T}{M_1}\right)^2  
  \right.
\nonumber\\
&& \hspace{2cm}
\left.  
  +\frac{7 \pi^2}{20}\,|\lambda_t|^2\,\left( \frac{T}{M_1}\right)^4
+ \left(  \frac{5}{6} \lambda_h - \frac{2g^2+g'^2}{24} \right) \frac{ k^2 \,T^2 }{M^4_1}
  \right] .
\label{finalRes}
\end{eqnarray}
This expression is accurate at fourth order in the Yukawa couplings, at order $M_1/M_i$ and 
at first order in the SM couplings, and, for each coupling, it provides the leading thermal correction. 
It also provides the leading thermal correction proportional to the three-momentum of $N_1$.

\begin{figure}[t!]
\centering
\includegraphics[scale=0.91]{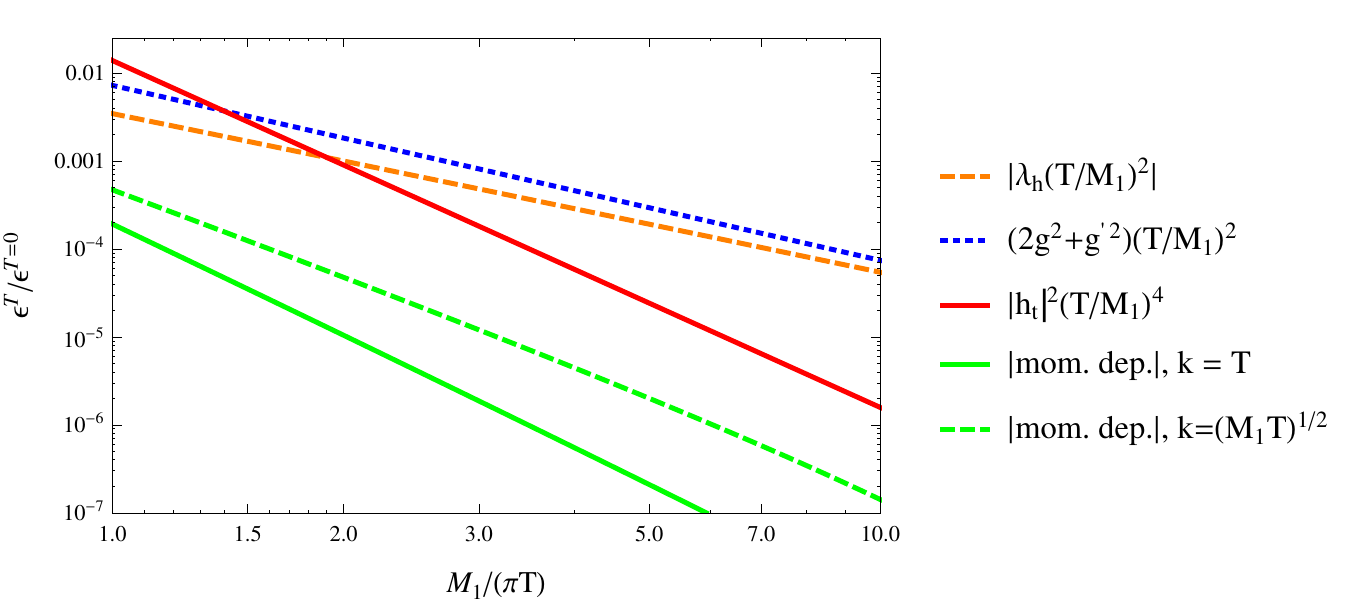}
\caption{
  Thermal corrections to the CP asymmetry of a RH neutrino decaying into leptons and antileptons as a function of the temperature
  in the hierarchical case\cite{Biondini:2016arl}.
The orange dashed line shows the contribution proportional to the Higgs self-coupling (the sign of the contribution 
has been changed to make it positive), the blue dotted line shows the contribution proportional to the gauge couplings 
and the red continuous line shows the contribution proportional to the top Yukawa coupling. 
The green lines show the leading thermal contribution proportional to the neutrino momentum 
(also in this case the sign of the contribution has been changed to make it positive).
For the green continuous line, we take the neutrino momentum to be $T$, whereas, for the green dashed line, we take it to be $\sqrt{M_1T}$.
The SM couplings have been computed at the scale $\pi T$ with one-loop running. 
The different thermal contributions to the CP asymmetry have been normalized with respect to the zero-temperature result expanded at leading order in $M_1/M_i$.
The neutrino mass has been taken $M_1 = 10^7$~GeV.}
\label{fig_plotCPcontributions}
\end{figure}

The leading thermal corrections proportional to the Higgs self-coupling $\lambda_h$ and to the gauge couplings $g$ and $g'$ 
are of relative order $(T/M_1)^2$, 
whereas those proportional to the top Yukawa coupling $|h_t|^2$ are of relative order $(T/M_1)^4$.
We show the different contributions in \fref{fig_plotCPcontributions}. 
At low temperatures, thermal corrections proportional to the Higgs self-coupling and to the gauge couplings dominate,
whereas at temperatures closer to the neutrino mass, the suppression in $T/M_1$ becomes less important 
and the  most numerically relevant corrections turn out to be those proportional to the top Yukawa coupling.
In \fref{fig_plotCPcontributions}, we also show the thermal contribution to the CP asymmetry due to a moving Majorana neutrino, 
which is of relative order $ k^2 \,T^2/M^4_1$. We plot this contribution for the case of a neutrino with momentum $T$ 
and for the case of a neutrino in thermal equilibrium with momentum $\sqrt{M_1T}$. 
We see that, for the considered momenta, the effect of a moving neutrino on the thermal CP asymmetry is tiny. 
In general, thermal effects are small in the hierarchical case, being at most of the order of a few percent for temperatures around the lightest neutrino mass. 
CP asymmetry and thermal effects may get enhanced if the two lightest neutrinos have almost degenerate masses.

Thermal corrections to the CP asymmetry in the case of two almost degenerate RH neutrinos
with masses $M_1$ and $M_2$, and $0 < \Delta = M_2-M_1 \ll  M_1$, were computed in Ref.~\cite{Biondini:2015gyw}.
In the nearly degenerate case, it makes sense to distinguish between direct and indirect contributions.
It is because of the indirect contributions, which can be resonant, that the CP asymmetry and its thermal corrections
may become enhanced in the nearly degenerate case.
The thermal corrections to the direct CP asymmetries for the two RH neutrinos $N_1$ and $N_2$ at rest read 
\begin{eqnarray}
  \epsilon^{T,{\rm direct}}_{1\alpha} &=& \frac{{\rm{Im}}\left[ (\lambda^{*}_1\lambda_2)(\lambda^{*}_{\alpha 1}\lambda_{\alpha 2})\right] }{8 \pi |\lambda_1|^2}
  \left(  \frac{T}{M_1} \right)^2 
\left\lbrace   \lambda_h \left[ 2-\ln 2+\left( 1-3\ln 2 \right) \frac{\Delta}{M_1}\right]  \right.
\nonumber \\
&& 
\left.
- \frac{g^2}{16}\left[ 2- \ln2 +\left( 3 - 5 \ln 2\right) \frac{\Delta}{M_1}  \right]  
- \frac{g'^2}{48}\left[ 4- \ln2 +\left( 1 - 5 \ln 2\right) \frac{\Delta}{M_1}  \right]  \right\rbrace,
\nonumber\\
\label{CPnu1_flavor}
\\
\epsilon^{T,{\rm direct}}_{2\alpha} &=& -\frac{{\rm{Im}}\left[ (\lambda^{*}_1\lambda_2) (\lambda^{*}_{\alpha 1}\lambda_{\alpha 2}) \right] }{8 \pi |\lambda_2|^2}
\left(  \frac{T}{M_1} \right)^2 
\left\lbrace   \lambda_h \left[ 2-\ln 2-\left( 9 - 5\ln 2 \right) \frac{\Delta}{M_1}\right]  \right.
\nonumber \\
&& 
\left.
- \frac{g^2}{16}\left[ 2- \ln2 - 7 \left( 1 - \ln 2\right) \frac{\Delta}{M_1}  \right]  
- \frac{g'^2}{48}\left[ 4- \ln2 -\left( 9 - 7 \ln 2\right) \frac{\Delta}{M_1}  \right]  \right\rbrace
\nonumber\\
&& + \frac{{\rm{Im}}\left[ (\lambda^{*}_2 \lambda_1)(\lambda^{*}_{\alpha 1}\lambda_{\alpha 2}) \right] }{2 \pi  |\lambda_2|^2}
\left(  \frac{T}{M_1} \right)^2\lambda_h \frac{\Delta}{M_1}, 
\label{CPnu2_flavor}
\end{eqnarray}
which are accurate at fourth-order in the Yukawa couplings, at first order in the SM couplings, at order $\Delta/M_1$ and at order $(T/M_1)^2$.
The leading thermal corrections to the indirect CP asymmetries for the two Majorana neutrinos at rest read
\begin{eqnarray}
\epsilon_{1\alpha}^{T,{\rm indirect}} &=& -\frac{\epsilon_{1\alpha}^{T=0,{\rm indirect}}}{3} \,\left( |\lambda_2|^2- |\lambda_1|^2 \right)\,
\frac{M_1}{\Delta}\,\frac{T^2}{M^2_1} ,
\label{indirect1T_fla} \\
\epsilon_{2\alpha}^{T,{\rm indirect}} &=& -\frac{\epsilon_{2\alpha}^{T=0,{\rm indirect}}}{3} \,\left( |\lambda_2|^2- |\lambda_1|^2 \right)\,
\frac{M_1}{\Delta}\,\frac{T^2}{M^2_1}, 
\label{indirect2T_fla}
\end{eqnarray}
which are accurate at zeroth-order in the SM couplings and at leading order in $\Delta/M_1$.
The $T=0$ CP asymmetry $\epsilon_{1\alpha}^{T=0,{\rm indirect}}$ may be read from \eref{CPself} for $i=2$, whereas $\epsilon_{2\alpha}^{T=0,{\rm indirect}}$ follows from $\epsilon_{1\alpha}^{T=0,{\rm indirect}}$
after the change $1 \leftrightarrow 2$.

The Yukawa coupling combination ${\rm{Im}}\left[ (\lambda^{\dagger}\lambda)_{21}(\lambda^{*}_{\alpha 1}\lambda_{\alpha 2}) \right]$ in \eref{CPnu2_flavor}
is absent in the expression of the direct CP asymmetry for the neutrino of type 1 in \eref{CPnu1_flavor}.
The origin of this contribution can be traced back to the kinematically allowed transition $N_2 \to N_1$ when $M_2 > M_1$,
which provides an additional source of CP asymmetry.

For what concerns the indirect CP asymmetry, we see that both the $T=0$ contribution in \eref{CPself} and the thermal contributions
in \eref{indirect1T_fla} and \eref{indirect2T_fla} become large for $\Delta$ close to zero, i.e.~in the nearly degenerate limit.
For mass differences $\Delta$ comparable with the widths $\Gamma_k$ of the Majorana neutrinos, these should be resummed in the neutrino propagators.
The resummation of the widths amounts to the replacement
\begin{equation}
  \frac{1}{\Delta} \to  \frac{\Delta}{\Delta^2 + (\Gamma_{2} - \Gamma_{1})^2/4},
 \label{res_replacement_EFT}
\end{equation}
in the $T=0$ and thermal expressions of the CP asymmetries (explicit expressions can be found in Ref.~\cite{Biondini:2015gyw}). 
At leading order, the RH-neutrino widths stem from the decay into a lepton and a Higgs boson and are $\Gamma_k \approx M_k |\lambda_{k}|^2/(8\pi)$. To capture the saturation of the resonant enhancement for $\Delta \lesssim \Gamma_k$, the replacement
in \eref{res_replacement_EFT} is not sufficient. Instead, it is necessary to include also coherent transitions between the Majorana neutrino
states, as described in detail in chapter \cite{leptogenesis:A03} of this review.
\subsection{CP-violating parameter and lepton asymmetry with non-equilibrium quantum field theory}\label{Ch4:sec3.2}
Within non-equilibrium field theory, an equation of motion for the lepton asymmetry can be obtained starting from the lepton density,
which is related to the temporal component of the lepton number current
\begin{equation}
  n_L = \frac{1}{V}\int_V d^3\bm{x} \langle J^0(t,\bm{x}) \rangle \,, \quad J^\mu \equiv \sum_{\alpha=e,\mu,\tau}\bar \ell_\alpha(x) \gamma^\mu \ell_\alpha(x)\,.
\end{equation}
Its time evolution, using the Schwinger-Dyson equation for the lepton two-point function
$S_{\alpha\beta}(x,y)=\langle T_{\cal C} \ell_\alpha(x)\bar\ell_\beta(y)\rangle$, is given by \cite{Buchmuller:2000nd,DeSimone:2007gkc,Garny:2009rv,Beneke:2010wd,Anisimov:2010dk,Garny:2011hg}
\begin{equation}
  (\partial_t +3H)n_L  =  - \frac{1}{V} \int d^3\bm{x} \int_{\cal C} d^4y \, \mathrm{tr}\left[ \Sigma_{\alpha\beta}(x,y) S_{\beta\alpha}(y,x) - S_{\alpha\beta}(x,y) \Sigma_{\beta\alpha}(y,x) \right] \;.
  \label{kb_lep}
\end{equation}
Here, $x_0=t$, $T_{\cal C}$ indicates time-ordering along the closed-time path $\cal C$, and $y^0$ is integrated over $\cal C$, $\Sigma_{\beta\alpha}$ is the lepton self-energy.
For a nearly CP-symmetric state, the right-hand side can be expanded in the lepton chemical potential, $\partial_t n_L  = S - W + \dots$,
where the zeroth-order contribution  is the source term $S$ and the first order is the washout term $W$.
\begin{figure}[t]
\begin{center}
\includegraphics[scale=0.48]{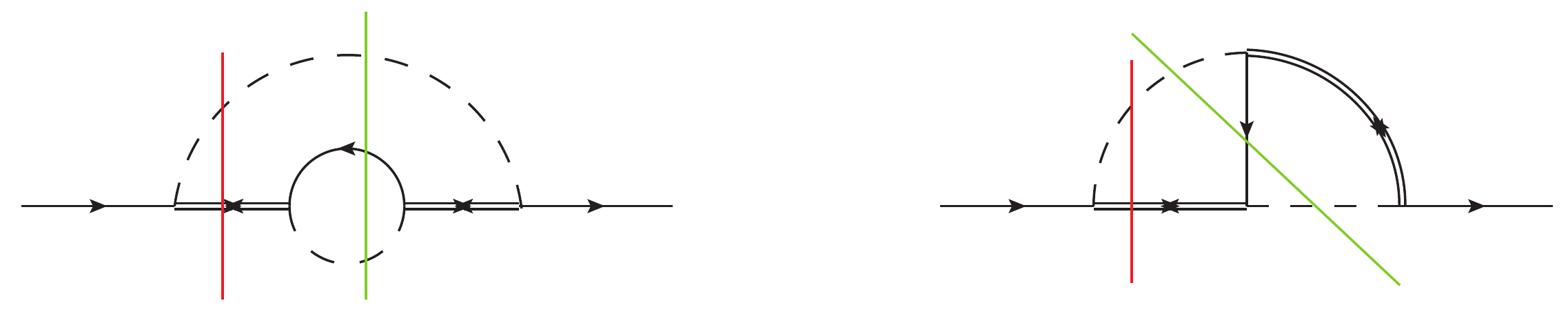}
\end{center}
\caption{\label{MG:selfenergy} Two-loop contributions to the lepton self-energy at ${\cal O}(\lambda^4)$. The red and green lines illustrate cuts that correspond to
decay (red) and scattering (green) contributions within the Boltzmann approach.}
\end{figure}
The leading source term can be obtained by inserting the two-loop lepton self-energy at order ${\cal \lambda}^4$, shown in \fref{MG:selfenergy}, and expanding all propagators around a CP-symmetric
state\cite{Garny:2011hg}. Furthermore, when using a Kadanoff-Baym ansatz for all propagators\cite{Frossard:2012pc}, and assuming that only $N_1$ deviates from equilibrium and that
$S_{\alpha\beta} \propto \delta_{\alpha\beta}$ (unflavored regime), one obtains for times $t\gg 1/M_1, 1/T$\cite{Garny:2010nj,Beneke:2010wd,Anisimov:2010dk,Frossard:2012pc}
and up to terms ${\cal O}(e^{-M_i/T})$ (see Ref.~\cite{Garbrecht:2010sz} for a discussion of these
contributions)
\begin{equation}\label{MG:source}
S = 4|\lambda_1|^2\int_{kpq} (2\pi)^4\delta(p-k-q) \, p\cdot k\, \epsilon_{1}(p, k, q) \, \delta f_{N\bm{p}}(t) (1-f^{{\rm{eq}}}_\ell(k)+f^{{\rm{eq}}}_\phi(q)) \;, 
\end{equation}
where $ \delta f_{N\bm{p}}(t) = f_N(p,t)-f_N^{{\rm{eq}}}(p)$ denotes the deviation of the $N_1$-distribution from thermal equilibrium, 
$\int_{k} = \int \frac{d^3 \bm{k}}{2\omega_k(2\pi)^3}$,  $f^{{\rm{eq}}}_\ell(k)=1/(e^{\beta \omega_k}+1)$, $f^{{\rm{eq}}}_\phi(q)=1/(e^{\beta \omega_q}-1)$,
and
\begin{equation}
  \epsilon_1(p, k, q)  = \frac{{\rm{Im}}[(\lambda_1^* \lambda_i )^2]}{8\pi (\lambda^\dag\lambda)_{11}} 
                     \left( \frac{M_1M_i}{M_1^2-M_i^2}\frac{k^\mu L_\mu(p)}{p\cdot k}
                      - \frac12 \frac{M_1}{M_i}\frac{k^\mu K_{\mu i}(p,q)}{p\cdot k}\right)\;,
\label{CP_garny}
\end{equation}
where
\begin{eqnarray}
  L_\mu(p) &\equiv& 16\pi\int_{k'q'} (2\pi)^4\delta(p-k'-q') k'_\mu (1-f^{{\rm{eq}}}_\ell(k')+f^{{\rm{eq}}}_\phi(q')) \,, \nonumber\\
  K_{\mu i}(p,q) &\equiv& 16\pi\int_{k'q'} (2\pi)^4\delta(p-k'-q') k'_\mu (1-f^{{\rm{eq}}}_\ell(k')+f^{{\rm{eq}}}_\phi(q')) \frac{M_i^2}{M_i^2-(q-k')^2}\,. \nonumber
\end{eqnarray}
Let us stress that the unflavored CP asymmetry in \eref{CP_garny} comprises thermal effects, which are included through the thermal distribution functions in $ L_\mu(p)$ and $K_{\mu i}(p,q)$. 
The two terms in $\epsilon_1$ arise from the two diagrams in \fref{MG:selfenergy} and
correspond to wave- and vertex-contributions, respectively.
At finite temperature, the integral $L_\mu$ can be expressed as \cite{Beneke:2010wd,Frossard:2012pc}
\begin{eqnarray}
  L_0 &=& \frac{2T}{y}I_1(y_0,y) \,, \\
  L_i &=& \frac{2p_i}{y^3}\left(y_0I_1(y_0,y)-\frac12(y_0^2-y^2)(1+x_\ell-x_\phi)I_0(y_0,y)\right)\,,
\end{eqnarray}
where $y_0=p_0/T$, $y=|p|/T$, $x_{\ell(\phi)}=m_{\ell(\phi)}/T$, and
\begin{equation}
  I_n(y_0,y) \equiv \int_{z_-}^{z_+} dz \, z^n \left( 1+\frac{1}{e^{y_0-z}-1} - \frac{1}{e^z+1}\right)\,,
\end{equation}
with $z_\pm = \left[ y_0(1+x_\ell-x_\phi)\pm y\lambda^{\frac12}(1,x_\ell,x_\phi)\right] /2$.
For $T, m_{\ell}, m_{\phi}\to 0$, the integrals are given by $L_\mu(p)\to p_\mu$ and $K_{\mu i}(p,q)\to -2p_\mu x_i (1-(1+x_i)\ln(1+1/x_i))$, 
where $x_i=M_i^2/p^2$ and $p^2=M_1^2$. For $M_i\gg M_1$, both integrals agree, i.e.~$K_{\mu i}(p,q)\to L_\mu(p)$.
Explicit analytic results for $I_n$ in the massless limit $m_\ell=m_\phi=0$ can be found in Ref.\cite{Beneke:2010wd}.
In the zero-temperature limit, the CP-violating parameter approaches the usual vacuum expression, i.e.~when adding  \eref{CPvert} and \eref{CPself} and summing over the lepton flavor $\alpha$. 
In this limit, the source term in \eref{MG:source} agrees with the conventional Boltzmann result (cf. \sref{basics_lep}).
At finite temperature, the source term in \eref{MG:source} can equivalently be written in the form
\begin{eqnarray}
S &=& 8 {\rm{Im}}[(\lambda_1^* \lambda_i )^2] \int_{pkqk'q'} \!\!\!\!\!\!\!
  (2\pi)^4\delta(p-k-q) (2\pi)^4\delta(p-k'-q') \, k^\mu k'_\mu M_1M_i \delta f_{N \bm{p}}(t) \nonumber\\
  & \times & \left( \frac{1}{M_1^2-M_i^2} - \frac12 \frac{1}{M_i^2-(q-k')^2}\right)   (1-f^{{\rm{eq}}}_\ell(k)+f^{{\rm{eq}}}_\phi(q))  (1-f^{{\rm{eq}}}_\ell(k')+f^{{\rm{eq}}}_\phi(q'))  \;. \nonumber\\
\end{eqnarray}
This form allows for a direct comparison to the Boltzmann result:
the first factor $(1-f^{{\rm{eq}}}_\ell(k)+f^{{\rm{eq}}}_\phi(q))$ originates from the quantum statistical terms in the Boltzmann equation, while the second factor
$(1-f^{{\rm{eq}}}_\ell(k')+f^{{\rm{eq}}}_\phi(q'))$ would be absent in the usual Boltzmann treatment and takes into account thermal corrections to the CP-violating loop amplitude. 
A notable property is that the source term is actually symmetric with respect to the interchange $q,k\leftrightarrow q',k'$ (using that $q-k'=q'-k$).
This can be understood within the closed-time path approach: the source term originates from closing the lepton self-energies shown in \fref{MG:selfenergy},
with the external lines connected by a lepton propagator. In this diagram, both lepton/Higgs lines with momenta $q,k$ and $q',k'$, respectively, appear symmetrically.\footnote{
This can lead to cancellations in scenarios where the CP asymmetry is a purely thermal effect, for example in the context of
soft leptogenesis \cite{Garbrecht:2013iga}.}

Furthermore, within the Boltzmann approach, it is necessary to rely on the so-called RIS subtraction procedure to obtain a consistent
source term (see \sref{basics_lep}), which can be cumbersome when including quantum-statistical terms and thermal effects\cite{Giudice:2003jh,Frossard:2012pc,Dev:2014laa}. 
This is not the case within the closed-time path formalism, where consistency comes ``for free''. The correspondence between the closed-time path formalism
and the Boltzmann approach can be illustrated by considering all possible cuts of the self-energies, which are indicated by the colored lines
in \fref{MG:selfenergy}. 

The source term can be further simplified in the strongly hierarchical limit $M_j\gg M_1$:
\begin{eqnarray}
S &=& \frac{|\lambda_1|^2}{4\pi } \, \epsilon_1^{T=0}\, \int_p L^\mu(p)L_\mu(p)\delta f_{N \bm{p}}(t)\,.
\end{eqnarray}
Quantitatively, including the finite-temperature corrections leads to an enhancement of the CP asymmetry that is exponentially suppressed for $T\ll M_1$
and numerically of ${\cal O}(1)$ for $T\sim M_1$. When naively taking into account thermal masses for the lepton and Higgs, the enhancement of the CP asymmetry
becomes smaller again for even larger temperatures. When $T\sim M_1/g$, the source term discussed above vanishes, and one expects a resummation similar to the
one described for $N_1$-production to become necessary to capture the leading-order result.

\section{Different approaches to rate equations for particle dynamics in the early Universe}\label{Ch4:sec4}
In this section, the evolution equations for leptogenesis are addressed. These equations are needed to obtain the time evolution of RH neutrino and lepton-asymmetry number densities. First, we discuss how the Boltzmann equations (BEs) can be derived from a full quantum mechanical set of equations and different methods to solve the latter. Afterwards, we move to effective kinetic equations, valid to all orders in the SM interactions and obtained by the main assumption of a large separation between the time scales of leptogenesis. NLO corrections to the rates are easier to handle in this second approach.
\subsection{On the applicability and limitation of standard Boltzmann equations}\label{Ch4:sec4.1}
As discussed in \sref{basics_lep}, the traditional approach to understanding the kinematics of leptogenesis is based on BEs. 
The underlying physical picture is one of classical particles that propagate freely between occasional scattering processes.  It is however clear that this approach has to be generalized and improved:  a net lepton number generation is a pure quantum
phenomenon, which takes place in the hot
early Universe.  Moreover, the Boltzmann equations suffer from the so-called double
counting problem, which can be solved only in some limiting cases. These issues have motivated more careful analyses in recent years: how are particles with finite lifetime properly accommodated in the Boltzmann picture, which assumes asymptotic initial and final states? Can the (quasi-)particles be defined consistently in a medium at finite density? Is the quantum interference that makes the difference between decays into particles and antiparticles affected by the medium?

Questions like these need to be addressed on a case-by-case basis for the various conceivable leptogenesis scenarios. To this end, the derivation has to start from a consistent set of quantum equations whose applicability is not in question. For the late-time limit of hierarchical thermal leptogenesis, these considerations have led to slight modifications\cite{Beneke:2010wd,Anisimov:2010dk,Garny:2011hg,Frossard:2012pc,Garny:2010nj,Garbrecht:2010sz} but otherwise put the Boltzmann approach on a sounder basis. Since, without strong assumptions, the proper quantum equations are hard to solve directly, there is also a great interest in obtaining generalized quantum kinetic equations simple enough for phenomenological analyses. These may be of a Boltzmann-like form, even if the physical picture of freely propagating quasi-particles is not applicable (i.e.~Boltzmann-type equations obtained under different or weaker assumptions).

A suitable starting point are self-consistent Schwinger-Dyson equations or an equivalent set of Kadanoff-Baym equations (KBEs)\cite{Baym:1961zz}, which are full equations of motion for out-of-equilibrium quantum systems. They are normally obtained via the Schwinger-Keldysh formalism\cite{Keldysh:1964ud}. The basic objects under study are the expectation values of the two-point functions
of the fields with the time argument belonging to the closed-time path. As far as leptogenesis is concerned, Higgs bosons, leptons/antileptons and Majorana neutrinos are the most important. Let us call the generic two-point function $G(x,y)$, which obeys a  Schwinger-Dyson equation $G^{-1}(x,y)=G^{-1}_{0}(x,y) - \Pi(x,y)$, where the free two-point function $G_0$ and self-energy $\Pi$ enter. KBEs are integro-differential equations that relate the physically relevant quantities of the out-of-equilibrium quantum fields with the imaginary and real parts of the retarded and advanced self-energies of the system.\footnote{In out-of-equilibrium dynamics, the Kubo-Martin-Schwinger relation is not satisfied so the system cannot be described by just one propagator. A stringent derivation of BEs will necessarily involve a definition that relates those quantum objects to distribution functions or number densities.} If the self-energies are obtained in a consistent manner (2PI effective action/Phi-derivable approximation), KBEs are known to be a suitable starting point for the derivation of {\em conserving} quantum kinetic equations. It is useful to express the two-point function $G(x,y)$ in terms of the spectral function and statistical function, $G^{-}(x,y)$ and $G^{+}(x,y)$ respectively.\footnote{The definition of the spectral and statistical correlator is $G^{-}(x,y) \equiv i \langle \left[ \phi(x),\phi(y) \right] \rangle $ and $G^{+}(x,y) \equiv \langle \left\lbrace  \phi(x),\phi(y) \right\rbrace \rangle /2 $ for a scalar field, see e.g.\cite{Drewes:2012qw}.} An example of a KBE is written in \eref{kb_lep} in \sref{Ch4:sec3.2}.   

These equations are hard to assess directly but may be solved through different sets of approximations. If one insists on obtaining an evolution equation for a one-particle distribution function, its definition in terms of the quantum objects of the KBEs is one of several necessary steps. In typical leptogenesis scenarios, the uncertainties are mostly in the solution of the KBEs for the RH neutrinos, whose evolution can depart far from equilibrium and whose mixing can be relevant for the generation of the asymmetry. Since the quantitatively-controlled first-principles derivation is key to answering the questions stated above, we sketch two known paths that lead to Boltzmann-type equations in the following. A complete analysis will usually include numerical estimates of the accuracy of the various approximations made.

First, we discuss the Fourier-space method, which relies on the separation of the slow macroscopic evolution of mean coordinates and the fast microscopic evolution\cite{Berges:2002wt,Prokopec:2003pj,Garny:2010nz,Frossard:2012pc}. It requires the transformation to the Wigner space coordinates $(x,y) \to (X,s)$, where ${\bf s}={\bf x}-{\bf y}$, $\Delta t=x_0-y_0$, ${\bf X}=({\bf x}+{\bf y})/2$, $t=(x_0+y_0)/2$, where ${\bf s}$ and $\Delta t$ are related to the microscopic scales ($\Delta t \sim 1/M_k$), and ${\bf X}$ and $t$ to the macroscopic scales ($t\sim 1/\Gamma_k,1/H$).  Consequently a Fourier transform is performed in the microscopic variables ${\bf s}$ and $\Delta t$, which give the momentum of the out-of-equilibrium field ${\bf k}$ and a frequency $\omega$, which can be related to the energy at later times. The so-obtained equations are still exact but not useful in practice. 
The next step consists of a {\em gradient expansion} in $t$ by exploiting the small changes in the mean coordinates. If the system is sufficiently close to equilibrium, one has for the relevant contributions $\Delta t\ll t$ and, since a homogeneous and isotropic Universe is assumed,  one may drop ${\bf X}$. The solutions obtained for the spectral function $G^{-}(x,y)$ incorporate both the thermal masses and a thermal width, traced back to the real and imaginary parts of the self-energies entering the corresponding evolution equation.  
Keeping only first-order contributions, quantum-kinetic equations that generalize the Boltzmann equations can be obtained for the statistical propagator. 

One may progress further in this direction by assuming a proportional relationship between the statistical propagator and the spectral function that generalizes their equilibrium relationship (Kadanoff-Baym ansatz). Using an approximate quasi-particle solution for the spectral function, which is proportional to $\delta (p^2 -m^2)$, puts the momentum on the mass shell. This latter step allows the integration of the energy coordinate of the four-momenta. Its accuracy should be checked by comparing with finite-width results. If multiple on-shell peaks happen to be close-by and overlap, such as in resonant leptogenesis, it is not necessarily good (see the companion chapter\cite{leptogenesis:A03}). Otherwise it defines one-particle distribution functions and the Boltzmann-type equations that govern their evolution.

A second approach makes use of the Wentzel-Kramers-Brillouin (WKB) approximation\cite{Wentzel:1926aor,Kramers:1926njj,Brillouin:1926blg}. In contrast to the gradient expansion, the WKB method does not rely on a Fourier transform in relative times (also known as the ``two-time formalism''\cite{Drewes:2012qw}). It also works far out-of-equilibrium and does not rely on an on-shell approximation or any other a priori assumption about the form of the correlation functions, such as the Kadanoff-Baym ansatz. 
The statistical propagator can be expressed in terms of a generalized distribution function that
follows a generalized BE, i.e.~a first-order differential equation that is local in time. The accuracy
of the BE is controlled by the accuracy of the WKB solution compared to the full KBE. The solutions found with this method are valid under
the very general physical assumptions of weak coupling and separation of macroscopic and microscopic time scales. 

The smallness of the parameters that control the accuracy of the WKB solution for $G^{\pm}$ is also required for the convergence of the gradient expansion in the Wigner-space approach. Hence, the applicability of both techniques is similar. The latter has a closer relationship to the diagrammatic S-matrix expansion in vacuum as both are performed in momentum space. If the finite width is taken into account, a full resummation is, however, required. On the other hand, the WKB method does not depend on such a resummation and treats this issue in a more intuitive manner.
\subsection{Effective kinetic equations and real-time correlation functions at finite temperature}\label{Ch4:sec4.2}
There are several circumstances that make leptogenesis relatively 
simple and allow for a relatively rigorous theoretical 
treatment: (i) most degrees of freedom are in thermal equilibrium,  because their
equilibration rate is much larger than the Hubble expansion rate $H$; 
(ii) there is a separation of time scales: the degrees of freedom which 
are out of equilibrium evolve much more slowly than most other degrees
of freedom and 
(iii) the system is  homogeneous in space.
The Hubble rate is typically of the same order as, or larger than the
equilibration rate of the slowly changing, or slow for short,
degrees of freedom.  Therefore, the fast degrees of freedom create a
quasi-equilibrium state, which is characterized by the temperature $ T
$ and the values of the slow ones.

In the absence of expansion the time derivative of the slow degrees of
freedom $ y _ a $ can only depend on the temperature and on the values
of the $ y _ a $. We choose the $ y _ a $ such that their expectation
values vanish in complete thermal equilibrium.  The typical time
scale on which they evolve is smaller than the inverse temperature.
The non-equilibrium state of the
system is then completely determined by the temperature, and the $ y _
a $.\footnote{It could also depend on the chemical potentials
of practically conserved charges, but we assume that these vanish.} 
If the $ y _ a $ are sufficiently small, one can expand to
linear order, so that, in the absence of expansion, the effective
kinetic equations take the form
\begin{equation}
  \dot y_{a  }=-\gamma_{a  b  }\, y_{b  }
  \label{eom} 
  .
\end{equation}
The real coefficients $ \gamma  _ { ab } $ only depend on the temperature.
One should stress that these equations do not require the validity of
any Boltzmann equation or quasiparticle picture. The only approximation
is based on the separation of time scales. 

What the slow variables are depends on the details of the model
parameters. As an example consider the simplest case of thermal
leptogenesis where (i) the masses of the RH neutrinos are
hierarchical, $ M _ 1 \ll M _ { i} $; (ii) the $ N _ 1 $-decay
rate is much larger than $ H $, which is the so-called strong washout
regime; (iii) the rates for charged-lepton Yukawa interactions are
much smaller than $ H $; (iv) the weak sphaleron rate is much larger
than $ H $.  The asymmetry is then generated when $ T $ is smaller
than $ M _ 1 $, and when there are only $ N _ 1 $'s present in the
plasma.  Furthermore, the RH neutrinos are non-relativistic.  To a
first approximation, one can neglect their motion completely. Then the
only slow variables are the $ N _ 1 $-number density $ n _ { N _ 1 }
$, and the baryon minus lepton number density $ n _ { B - L } $.\footnote{At very high $T$, one only has to account for a single lepton number,
which is the one which is produced in the $N_1$ decays. At lower
temperature, when charged-lepton Yukawa interactions are in
equilibrium, one has to account for different lepton flavors.}
Taking into account the Hubble expansion, the effective kinetic
equations (\ref{eom}) become\cite{Bodeker:2013qaa} 
\begin{align}
  D _ t n _ { N _ 1 } 
  & = 
  - \gamma  _ { N _ 1 } ( n _ { N _ 1 } -  n _ { N _ 1 }  
  ^ {\rm eq } ) - \gamma  _ { N _ 1, B - L } n _ { B - L } \, ,
  \label{eomN} 
  \\
  D _ t n _ { B - L } 
  & = 
  - \gamma  _ { B - L, N _ 1 } ( n _ { N _ 1 } -  n _ { N _ 1 }  
  ^ {\rm eq } ) - \gamma  _ {  B - L }  n _ { B - L } \, ,
  \label{eomasy} 
\end{align} 
where $ D _ t \equiv d / d t + 3 H $, and $
n _ { N _ 1 } ^ {\rm eq } $ is the equilibrium number density of $ N _
1 $.
\Eref{eomN} and \eref{eomasy} are valid to all orders in
SM couplings and on time scales much larger than the
equilibration time of the fast degrees of freedom.  

The coefficients $ \gamma _ { ab } $ in \eref{eom} can be
determined in thermal 
field theory by using the theory of quasi-stationary
fluctuations~\cite{Landau:1980mil}. Even in thermal equilibrium, the $
y _ a $ are not constant in time but fluctuate around their
equilibrium values.  Their thermal fluctuations can be described by the
same classical equations of motion as \eref{eom}, except that there
is an additional Gaussian white noise, which describes the effect of
fluctuations of the fast variables.  These equations can be used to
compute the real-time autocorrelation functions of the fluctuations of
$ y _ a $.
These correlation functions can also be computed in the underlying
microscopic quantum field theory. 
For $ \omega  \ll \omega_{{\rm UV}}  $, where $\omega_{{\rm UV}}$ is the
characteristic frequency of the spectator processes, the two results
have to match. In this way, one obtains the relation 
\begin{equation}
  \gamma_{ab}
  =
  T \omega \, \text{Im} \,\Pi  _{ac}^ {R}(\omega)
 \left(\chi^{-1}\right)_{cb} \, ,
 \qquad (  \gamma \ll\omega\ll\omega_{{\rm UV}} ) 
 \label{kubo} 
 .
\end{equation}
for the rates in \eref{eom}.\footnote{Here, the relation $\gamma \ll \omega $ has to be understood parametrically.
Please note that $\gamma$ need not be the neutrino production rate. \Eref{kubo} can also be used for the washout rate and for the asymmetry rate (and it has been used for that).} It is similar to the Kubo relations
for transport coefficients, and it contains  the retarded self-energy $\Pi_{ac}^ {R}$ of $ y _ a $ and $ y _ c $, as well as the real and symmetric matrix of susceptibilities 
$ \chi_{ab} \equiv \langle y_{a}y_{b}\rangle $.

Now one has to identify operators which represent the slowly changing
variables in the microscopic quantum field theory.
There are two different types of operators entering the susceptibilities  and the
retarded self-energies in \eref{kubo}: the
ones containing SM fields, and the ones containing only
the RH neutrinos.  The SM fields are in kinetic
equilibrium, and the corresponding operators are global charges which
would be conserved if certain Yukawa interactions are neglected,
\begin{equation}
  X_{a}
  =\int \! d^3x \, \bar{\ell }\gamma_{0}T^{\ell}_{a} \ell
  +\text{contributions from other fermions}
  \label{X} 
  .
\end{equation}
Here, $T^{\ell}_{a}$ is the generator of the corresponding symmetry
transformation acting on the left-handed lepton doublets $ \ell_\alpha $. 
The RH neutrinos interact only very weakly. Without any
interaction, the occupation number $ f _ { \bm{k} } $ of each field
mode with momentum $  \bm{k} $ would be conserved.  Assuming 
a homogeneous Universe, we do not need any variables
describing spatial variations.  For a free field, the spin-averaged
occupation number in a finite volume $ V $ is
\begin{equation}
  f_{\bm{k}} 
  =
  \frac 12 
  \sum_{s} c_{  \bm{k} s}^{\dagger}c _{ \bm{k} s}
  \label{f} 
  ,
\end{equation} 
with  creation and annihilation operators $c ^\dagger _{  \bm{k}
  s}$ and $c _{  \bm{k} s}$.
To include the Yukawa interaction 
of the RH neutrinos, as given in the Lagrangian (\ref{Lag1_full}),
we use the interaction picture.  Then the
sterile-neutrino 
field operator has the same form as in the free case, and we define the
occupation number in the interaction picture by
\eref{f}~\cite{Asaka:2006rw} If there are several flavors of
RH neutrinos present in the plasma, one has to work with a matrix
of occupation numbers (see also the accompanying chapter of this review\cite{leptogenesis:A01}). 

We expand the retarded two-point functions in \eref{kubo} in the
very small Yukawa couplings $ \lambda $ and keep only the leading
order.  They can be then factored out and no longer know about the
scale $ \gamma$, so that the restriction $ \omega \gg \gamma $ in the
matching formula (\ref{kubo}) can be dropped, and we can take the
limit $ \omega \to 0 $.  Furthermore, one can integrate out the
RH neutrinos treating them as free fields.  One is left with two-
or four-point functions of composite operators made of SM
fields. Consider, e.g., the washout rate $ \gamma _ { ab } $ in the
effective kinetic equation
\begin{align}
  \dot X _ a = - \gamma  _ { ab } X  _ b + \cdots 
  . 
\end{align}
If the small charged-lepton Yukawa interactions are neglected one
finds\footnote{Note that the product $ V (
  \chi ^ { -1 } ) _ { cb } $ is finite in the limit $ V \to {\infty }
  $.}\cite{Bodeker:2014hqa}
\begin{align} 
    \gamma  _ { ab }  = & \frac 12 T V \sum_{i} 
    \lambda  ^ \ast  _ {  \alpha i  } 
    \{ T _ a ^\ell , T _ c ^ \ell \}^{ } _ { \alpha  \beta   }
    \lambda   _ { \beta i  } 
   \left ( \chi  ^{ -1 } \right ) _ { cb } 
   \, \mathcal{W}(M_i)
  \label{pre_washout_rate} 
  .
\end{align}
Here, one sees the factorized structure; the function
\begin{align} 
  \mathcal{W}(M_i)
  & \equiv   -  
  \int \frac { d ^ 3 k } { ( 2 \pi  ) ^ 3 }
  \frac{ f'_{{\rm{eq}},N} ( \omega _ i )  } { 2 \omega _ i   }
  \, 
  {\rm Tr } \Big \{ 
   \slashed{ k } 
  \big [ 
    \rho  ( k )
 +  \rho   ( -k )\big ]
 \Big \} _ { k ^ 0 = \omega _ { i  } } \, ,
 \label{washout_rate}
\end{align} 
with $ \omega_{ i  } \equiv ( k^2 + M_i^2 ) ^ { 1/2 } $, 
does not depend on the sterile-neutrino Yukawa couplings $ \lambda _ {
  \alpha i }$. The prime on the Fermi-Dirac distribution for the RH neutrino denotes a derivative with respect to the energy and the fundamental object is a two-point function of the composite operators
of SM fields
\begin{align} 
   \rho   ( k ) 
   \, \equiv
   \frac 13
   \int \! {\rm d} ^ 4 x \,  
   e ^{ i k \cdot x } 
   \Bigl\langle \, 
      \Bigl\{ 
   (\phi^{c \dagger} \ell _ {  \alpha  }) ( x ) , 
   (\overline{ \ell\, } _ {\! \alpha    }\, \phi^c) ( 0 )  
      \Bigr\}
   \, \Bigr \rangle ^{ }_ 0
   \label{rho}
   .
\end{align} 
The trace in \eref{washout_rate}  is over Dirac spinor indices, and the
subscript 0 in \eref{rho} indicates that the thermal average is 
performed with $ \lambda  = 0 $. The factor $1/3$ is due to a flavor average.

The final step in computing the rates is to evaluate the correlation functions
of the SM fields, as well as the susceptibilities. 
At leading order in $ \lambda $, 
the susceptibilities of the sterile-neutrino occupation can be computed
in the free-field limit.
The susceptibilities of the charges in \eref{X} 
have been computed in Refs.~\cite{Bodeker:2014hqa,Bodeker:2015zda} to quadratic order in the SM
couplings. Interestingly, the leading correction to the free-field
result is parametrically of {\it linear} order due to infrared
effects. 

Recently,  the approach described in this section
has also been applied to the asymmetry rate $\gamma_{B-L,N_1} $~\cite{Bodeker:2017deo}. It has been computed at
NLO in SM couplings at $T=0$ in the hierarchical
limit, and the corrections were found to be of order 5\%. Thus, thermal
leptogenesis is theoretically well under control. Also
recently, the approach just described, based
on the separation of time scales and the resulting identification of
the slow modes, has been applied to leptogenesis at $T\simg M$ with RH neutrino masses below the electroweak scale~\cite{Ghiglieri:2017gjz}.  
A set of equations resembling the schematic
structure of Eqs.~(\ref{eomN}) and (\ref{eomasy}) has been derived in
the same spirit of this section, using a density matrix to keep track
of the slow sterile-neutrino degrees of freedom, i.e. momentum,
flavour and helicity, since it is found that the two helicity states
are produced and equilibrate at different rates in the presence of a
lepton asymmetry (see also Refs.\cite{Lello:2016rvl,Eijima:2017anv}).
\section{Conclusions and outlook}\label{Ch4:sec5}
If the baryon asymmetry in the Universe was generated via leptogenesis, the RH neutrinos experienced a
thermal medium of SM particles. It is a non-trivial task to include thermal effects to the different ingredients 
of the heavy-neutrino dynamics: production rate, CP asymmetries and washout rates. Those quantities enter the evolution equations for the
RH neutrino and lepton-asymmetry number densities. In order to include the effects of the SM heat bath systematically, NLO 
calculations at finite temperature are needed. 
Hence, thermal field theory is the tool to achieve a better knowledge of the key ingredients for leptogenesis.

The RH neutrino production rate is known at NLO (in-vacuum and thermal corrections) for all the kinematic regimes satisfying $M\simg \pi T$. A broad temperature range is covered, namely temperatures larger and smaller than the SM electroweak crossover $T_{c}\sim 160$ GeV. 
Different approaches have provided the same result for the non-relativistic regime, where NLO thermal corrections are power suppressed (see \eref{gamma_nonrel}). 

The status of the CP asymmetries
in RH neutrino decays is as follows. Thermal corrections at zeroth order in SM couplings are known and they are exponentially suppressed when $T \ll M$ (see \sref{Ch4:sec3.2}), whereas they can be of numerical relevance for temperatures closer to the neutrino mass. The NLO in-vacuum result has been recently carried out for a hierarchical heavy-neutrino
mass spectrum\cite{Bodeker:2017deo}. On the other hand, NLO thermal corrections have been studied in the framework of non-relativistic EFTs. For the hierarchical 
case, the leading thermal corrections, of relative order $(T/M_1)^2$, are known, whereas the subleading ones, of relative order $(T/M_1)^4$, comprise the contribution from the top-Yukawa coupling only at the time of this review. For the nearly degenerate case, thermal corrections of relative order $(T/M_1)^2$ have been derived so far. Such results are obtained in an unbroken phase of the SM.
Completing the thermal NLO result in the non-relativistic regime appears possible in the near future. Moreover, the derivation of the CP asymmetries in the relativistic regime is desirable, so as to cover a broader temperature range. A three-loop thermal calculation is what is needed to accomplish such a goal. 

Lepton-number washout rates come as a combination of the neutrino width and lepton-number susceptibilities. The latter are known at NLO and NNLO at $T > 160$ GeV. At leading order, numerical results are also available for $T < 160$ GeV.

The Boltzmann equations are the traditional approach to address the number-density evolution of RH neutrinos and lepton asymmetry. They rely on several approximations that have to be tested case-by-case for different leptogenesis scenarios. We have reviewed recent developments that allow for a rigorous derivation of Boltzmann equations from more fundamental quantum kinetic equations. The latter are not easy to solve and therefore approximation methods are needed (see \sref{Ch4:sec4.1}). Moreover, we discussed a complementary approach that relies only on the separation of time scales appearing in leptogenesis (see \sref{Ch4:sec4.2}). These effective kinetic equations are valid to all orders in the SM couplings. Indeed radiative corrections have already been successfully included in this framework. The neutrino production rate, CP asymmetries and washout rates are seen as the coefficients dictating the evolution of RH neutrino and lepton-asymmetry occupation numbers and they  are traced back to correlation functions at finite temperature. 

Finally, we comment on the different technical approaches discussed in this chapter. 
In \sref{Ch4:sec2.1}, the imaginary part of the retarded self-energy in the real-time formalism 
is considered, $\Pi_R= \Pi_{11} + \Pi_{12}$, which is related to the neutrino
thermal width. In the strict non-relativistic regime it suffices looking at 
the $\Pi_{11}$ component, since $\Pi_{12}$ is exponentially suppressed in 
the heavy-mass limit. Then the very same quantity is understood in terms of 
a spectral function in \sref{Ch4:sec2.1}, whose relation to the retarded self-energy 
reads $\rho = 2 \mathrm{Im}\,\Pi_R$
(actually all the correlation functions can be
expressed in terms of $\rho$ in thermal equilibrium)
and the imaginary-time formalism is adopted. There, one first computes
the Euclidean self-energy, which is then analytically continued 
into the retarded self-energy.
In \sref{Ch4:sec3.2}, we introduce the lepton asymmetry in the closed-time path formalism. 
Here, the expectation values are taken along the original Schwinger-Keldysh 
contour as for the real-time formalism. This formulation of thermal field theory 
is known to be better suited to address directly out-of-equilibrium dynamics, since
it corresponds to evolving the initial density operator in time along this contour. In the equilibrium
case, the vertical part of the real-time contour is indeed the initial Boltzmann
density operator. We further 
remark that the lepton asymmetry originates from the imaginary part of closed-time 
path self-energies. On the same footing, the imaginary part of correlation functions 
is the key ingredient for obtaining the coefficients of the evolution equations. As we saw 
in \sref{Ch4:sec4.2}, one ends up relating the imaginary part of retarded self-energies via 
a fluctuation-dissipation argument to the washout rate. The method sketched there
also includes the effect of the time evolution of the initial density operator, which
in this case factors the equilibrated fast modes from the out-of-equilibrium slow modes. 
The time evolution is governed by the equations of motions for the operators describing the slow
modes.

\section*{Acknowledgements}
We gratefully acknowledge the hospitality of the Munich Institute for Astro- and Particle Physics (MIAPP) of the DFG cluster of excellence ``Origin and Structure of the Universe'', where this work has been initiated.  The work of S.B. and M.L. was supported by the Swiss National Science
Foundation (SNF) under grant 200020-168988. N.B. and A.V. acknowledge support from the DFG cluster of excellence ``Origin and structure of the Universe" (www.universe-cluster.de). The work of S.M. was supported by FONDECYT 11130118. The work of P.M. was supported by STFC Grant No. ST/L000393/1 and a Leverhulme Trust Research Leadership Award. P.M. would like to thank Pasquale Di Bari and St\'{e}phane Lavignac for helpful discussions. A.S. acknowledges support from the grant 669668 -- NEO-NAT -- ERC-AdG-2014.

\bibliographystyle{ws-rv-van-mod2}

\bibliography{chapter-v2}

\begin{thebibliography}{75}
\providecommand{\natexlab}[1]{#1}
\providecommand{\url}[1]{\texttt{#1}}
\expandafter\ifx\csname urlstyle\endcsname\relax
  \providecommand{\doi}[1]{doi: #1}\else
  \providecommand{\doi}{doi: \begingroup \urlstyle{rm}\Url}\fi

\bibitem{Fukugita:1986hr}
M.~Fukugita and T.~Yanagida, {Baryogenesis Without Grand Unification},
  \emph{Phys. Lett.} {\bf B174}, \penalty0 45--47,  (1986).

\bibitem{Drewes:2013gca}
M.~Drewes, {The Phenomenology of Right Handed Neutrinos}, \emph{Int. J. Mod.
  Phys.} {\bf E22}, \penalty0 1330019,  (2013).

\bibitem{Kuzmin:1985mm}
V.~A. Kuzmin, V.~A. Rubakov, and M.~E. Shaposhnikov, {On the Anomalous
  Electroweak Baryon Number Nonconservation in the Early Universe}, \emph{Phys.
  Lett.} {\bf B155}, \penalty0 36,  (1985).

\bibitem{Buchmuller:2004nz}
W.~{Buchm\"uller}, P.~Di~Bari, and M.~{Pl\"umacher}, {Leptogenesis for
  pedestrians}, \emph{Annals Phys.} {\bf 315}, \penalty0 305--351,  (2005).

\bibitem{Davidson:2008bu}
S.~Davidson, E.~Nardi, and Y.~Nir, {Leptogenesis}, \emph{Phys. Rept.} {\bf
  466}, \penalty0 105--177,  (2008).

\bibitem{Buchmuller:1996pa}
W.~{Buchm\"uller} and M.~{Pl\"umacher}, {Baryon asymmetry and neutrino mixing},
  \emph{Phys. Lett.} {\bf B389}, \penalty0 73--77,  (1996).

\bibitem{Fogli:2011qn}
G.~L. Fogli, E.~Lisi, A.~Marrone, A.~Palazzo, and A.~M. Rotunno, {Evidence of
  $\theta_{13}>0$ from global neutrino data analysis}, \emph{Phys. Rev.} {\bf
  D84}, \penalty0 053007,  (2011).

\bibitem{Blanchet:2012bk}
S.~Blanchet and P.~Di~Bari, {The minimal scenario of leptogenesis}, \emph{New
  J. Phys.} {\bf 14}, \penalty0 125012,  (2012).

\bibitem{Pilaftsis:1998pd}
A.~Pilaftsis, {Heavy Majorana neutrinos and baryogenesis}, \emph{Int. J. Mod.
  Phys.} {\bf A14}, \penalty0 1811--1858,  (1999).

\bibitem{Fong:2013wr}
C.~S. Fong, E.~Nardi, and A.~Riotto, {Leptogenesis in the Universe}, \emph{Adv.
  High Energy Phys.} {\bf 2012}, \penalty0 158303,  (2012).

\bibitem{leptogenesis:A01}
P.~S.~B. Dev, P.~Di~Bari, B.~Garbrecht, S.~Lavignac, P.~Millington, and
  D.~Teresi, {Flavor effects in leptogenesis}, {\ttfamily arXiv:1711.02861},
  (2017).

\bibitem{Luty:1992un}
M.~A. Luty, {Baryogenesis via leptogenesis}, \emph{Phys. Rev.} {\bf D45},
  \penalty0 455--465,  (1992).

\bibitem{Kolb:1979qa}
E.~W. Kolb and S.~Wolfram, {Baryon Number Generation in the Early Universe},
  \emph{Nucl. Phys.} {\bf B172}, \penalty0 224,  (1980).
\newblock [Erratum: Nucl. Phys.B195,542(1982)].

\bibitem{Pilaftsis:2003gt}
A.~Pilaftsis and T.~E.~J. Underwood, {Resonant leptogenesis}, \emph{Nucl.
  Phys.} {\bf B692}, \penalty0 303--345,  (2004).

\bibitem{Fong:2010up}
C.~S. Fong and J.~Racker, {On fast CP violating interactions in leptogenesis},
  \emph{JCAP}. {\bf 1007}, \penalty0 001,  (2010).

\bibitem{Covi:1996wh}
L.~Covi, E.~Roulet, and F.~Vissani, {CP violating decays in leptogenesis
  scenarios}, \emph{Phys. Lett.} {\bf B384}, \penalty0 169--174,  (1996).

\bibitem{leptogenesis:A03}
P.~S.~B. Dev, M.~Garny, J.~Klaric, P.~Millington, and D.~Teresi, {Resonant
  enhancement in leptogenesis}, {\ttfamily arXiv:1711.02863},  (2017).

\bibitem{Salvio:2011sf}
A.~Salvio, P.~Lodone, and A.~Strumia, {Towards leptogenesis at NLO: the
  right-handed neutrino interaction rate}, \emph{JHEP}. {\bf 08}, \penalty0
  116,  (2011).

\bibitem{Laine:2011pq}
M.~Laine and Y.~Schroder, {Thermal right-handed neutrino production rate in the
  non-relativistic regime}, \emph{JHEP}. {\bf 02}, \penalty0 068,  (2012).

\bibitem{Barbieri:1999ma}
R.~Barbieri, P.~Creminelli, A.~Strumia, and N.~Tetradis, {Baryogenesis through
  leptogenesis}, \emph{Nucl. Phys.} {\bf B575}, \penalty0 61--77,  (2000).

\bibitem{Giudice:2003jh}
G.~F. Giudice, A.~Notari, M.~Raidal, A.~Riotto, and A.~Strumia, {Towards a
  complete theory of thermal leptogenesis in the SM and MSSM}, \emph{Nucl.
  Phys.} {\bf B685}, \penalty0 89--149,  (2004).

\bibitem{LeBellac}
M.~L. Bellac, Thermal field theory, \emph{Cambridge University Press}.  (1996).

\bibitem{Biondini:2013xua}
S.~Biondini, N.~Brambilla, M.~A. Escobedo, and A.~Vairo, {An effective field
  theory for non-relativistic Majorana neutrinos}, \emph{JHEP}. {\bf 12},
  \penalty0 028,  (2013).

\bibitem{Laine:2016hma}
M.~Laine and A.~Vuorinen, {Basics of Thermal Field Theory}, \emph{Lect. Notes
  Phys.} {\bf 925}, \penalty0 pp.1--281,  (2016).

\bibitem{Bodeker:2015exa}
D.~B{\"o}deker, M.~Sangel, and M.~Wormann, {Equilibration, particle production,
  and self-energy}, \emph{Phys. Rev.} {\bf D93}\penalty0 (4), \penalty0 045028,
   (2016).

\bibitem{Laine:2015kra}
M.~Laine and M.~Meyer, {Standard Model thermodynamics across the electroweak
  crossover}, \emph{JCAP}. {\bf 1507}\penalty0 (07), \penalty0 035,  (2015).

\bibitem{D'Onofrio:2015mpa}
M.~D'Onofrio and K.~Rummukainen, {Standard model cross-over on the lattice},
  \emph{Phys. Rev.} {\bf D93}\penalty0 (2), \penalty0 025003,  (2016).

\bibitem{D'Onofrio:2014kta}
M.~D'Onofrio, K.~Rummukainen, and A.~Tranberg, {Sphaleron Rate in the Minimal
  Standard Model}, \emph{Phys. Rev. Lett.} {\bf 113}\penalty0 (14), \penalty0
  141602,  (2014).

\bibitem{Garbrecht:2013gd}
B.~Garbrecht, F.~Glowna, and M.~Herranen, {Right-Handed Neutrino Production at
  Finite Temperature: Radiative Corrections, Soft and Collinear Divergences},
  \emph{JHEP}. {\bf 04}, \penalty0 099,  (2013).

\bibitem{Laine:2013vpa}
M.~Laine, {Thermal 2-loop master spectral function at finite momentum},
  \emph{JHEP}. {\bf 05}, \penalty0 083,  (2013).

\bibitem{Laine:2013lka}
M.~Laine, {Thermal right-handed neutrino production rate in the relativistic
  regime}, \emph{JHEP}. {\bf 08}, \penalty0 138,  (2013).

\bibitem{Besak:2010fb}
D.~Besak and D.~B{\"o}deker, {Hard Thermal Loops for Soft or Collinear External
  Momenta}, \emph{JHEP}. {\bf 05}, \penalty0 007,  (2010).

\bibitem{Anisimov:2010gy}
A.~Anisimov, D.~Besak, and D.~B{\"o}deker, {Thermal production of relativistic
  Majorana neutrinos: Strong enhancement by multiple soft scattering},
  \emph{JCAP}. {\bf 1103}, \penalty0 042,  (2011).

\bibitem{Besak:2012qm}
D.~Besak and D.~B{\"o}deker, {Thermal production of ultrarelativistic
  right-handed neutrinos: Complete leading-order results}, \emph{JCAP}. {\bf
  1203}, \penalty0 029,  (2012).

\bibitem{Ghiglieri:2016xye}
J.~Ghiglieri and M.~Laine, {Neutrino dynamics below the electroweak crossover},
  \emph{JCAP}. {\bf 1607}\penalty0 (07), \penalty0 015,  (2016).

\bibitem{Ghisoiu:2014ena}
I.~Ghisoiu and M.~Laine, {Right-handed neutrino production rate at $T > 160$
  GeV}, \emph{JCAP}. {\bf 1412}\penalty0 (12), \penalty0 032,  (2014).

\bibitem{Ghiglieri:2013gia}
J.~Ghiglieri, J.~Hong, A.~Kurkela, E.~Lu, G.~D. Moore, and D.~Teaney,
  {Next-to-leading order thermal photon production in a weakly coupled
  quark-gluon plasma}, \emph{JHEP}. {\bf 05}, \penalty0 010,  (2013).

\bibitem{Ghiglieri:2014kma}
J.~Ghiglieri and G.~D. Moore, {Low Mass Thermal Dilepton Production at NLO in a
  Weakly Coupled Quark-Gluon Plasma}, \emph{JHEP}. {\bf 12}, \penalty0 029,
  (2014).

\bibitem{Hernandez:2016kel}
P.~Hernandez, M.~Kekic, J.~Lopez-Pavon, J.~Racker, and J.~Salvado, {Testable
  Baryogenesis in Seesaw Models}, \emph{JHEP}. {\bf 08}, \penalty0 157,
  (2016).

\bibitem{Ghiglieri:2017gjz}
J.~Ghiglieri and M.~Laine, {GeV-scale hot sterile neutrino oscillations: a
  derivation of evolution equations}, \emph{JHEP}. {\bf 05}, \penalty0 132,
  (2017).

\bibitem{Shi:1998km}
X.-D. Shi and G.~M. Fuller, {A New dark matter candidate: Nonthermal sterile
  neutrinos}, \emph{Phys. Rev. Lett.} {\bf 82}, \penalty0 2832--2835,  (1999).

\bibitem{Ghiglieri:2015jua}
J.~Ghiglieri and M.~Laine, {Improved determination of sterile neutrino dark
  matter spectrum}, \emph{JHEP}. {\bf 11}, \penalty0 171,  (2015).

\bibitem{Venumadhav:2015pla}
T.~Venumadhav, F.-Y. Cyr-Racine, K.~N. Abazajian, and C.~M. Hirata, {Sterile
  neutrino dark matter: Weak interactions in the strong coupling epoch},
  \emph{Phys. Rev.} {\bf D94}\penalty0 (4), \penalty0 043515,  (2016).

\bibitem{Eijima:2017anv}
S.~Eijima and M.~Shaposhnikov, {Fermion number violating effects in low scale
  leptogenesis}, \emph{Phys. Lett.} {\bf B771}, \penalty0 288--296,  (2017).

\bibitem{Lello:2016rvl}
L.~Lello, D.~Boyanovsky, and R.~D. Pisarski, {Production of heavy sterile
  neutrinos from vector boson decay at electroweak temperatures}, \emph{Phys.
  Rev.} {\bf D95}\penalty0 (4), \penalty0 043524,  (2017).

\bibitem{Bodeker:2014hqa}
D.~B{\"o}deker and M.~Laine, {Kubo relations and radiative corrections for
  lepton number washout}, \emph{JCAP}. {\bf 1405}, \penalty0 041,  (2014).

\bibitem{Bodeker:2015zda}
D.~B{\"o}deker and M.~Sangel, {Order $g^2$ susceptibilities in the symmetric
  phase of the Standard Model}, \emph{JCAP}. {\bf 1504}\penalty0 (04),
  \penalty0 040,  (2015).

\bibitem{Eijima:2017cxr}
S.~Eijima, M.~Shaposhnikov, and I.~Timiryasov, {Freeze-out of baryon number in
  low-scale leptogenesis}, {\ttfamily arXiv:1709.07834},  (2017).

\bibitem{Covi:1997dr}
L.~Covi, N.~Rius, E.~Roulet, and F.~Vissani, {Finite temperature effects on CP
  violating asymmetries}, \emph{Phys. Rev.} {\bf D57}, \penalty0 93--99,
  (1998).

\bibitem{Bodeker:2013qaa}
D.~B{\"o}deker and M.~Wörmann, {Non-relativistic leptogenesis}, \emph{JCAP}.
  {\bf 1402}, \penalty0 016,  (2014).

\bibitem{Biondini:2016arl}
S.~Biondini, N.~Brambilla, and A.~Vairo, {CP asymmetry in heavy Majorana
  neutrino decays at finite temperature: the hierarchical case}, \emph{JHEP}.
  {\bf 09}, \penalty0 126,  (2016).

\bibitem{Bodeker:2017deo}
D.~B{\"o}deker and M.~Sangel, {Lepton asymmetry rate from quantum field theory:
  NLO in the hierarchical limit}, \emph{JCAP}. {\bf 1706}\penalty0 (06),
  \penalty0 052,  (2017).

\bibitem{Biondini:2015gyw}
S.~Biondini, N.~Brambilla, M.~A. Escobedo, and A.~Vairo, {CP asymmetry in heavy
  Majorana neutrino decays at finite temperature: the nearly degenerate case},
  \emph{JHEP}. {\bf 03}, \penalty0 191,  (2016).
\newblock [Erratum: JHEP08,072(2016)].

\bibitem{Buchmuller:2000nd}
W.~{Buchm\"uller} and S.~Fredenhagen, {Quantum mechanics of baryogenesis},
  \emph{Phys. Lett.} {\bf B483}, \penalty0 217--224,  (2000).

\bibitem{DeSimone:2007gkc}
A.~De~Simone and A.~Riotto, {Quantum Boltzmann Equations and Leptogenesis},
  \emph{JCAP}. {\bf 0708}, \penalty0 002,  (2007).

\bibitem{Garny:2009rv}
M.~Garny, A.~Hohenegger, A.~Kartavtsev, and M.~Lindner, {Systematic approach to
  leptogenesis in nonequilibrium QFT: Vertex contribution to the CP-violating
  parameter}, \emph{Phys. Rev.} {\bf D80}, \penalty0 125027,  (2009).

\bibitem{Beneke:2010wd}
M.~Beneke, B.~Garbrecht, M.~Herranen, and P.~Schwaller, {Finite Number Density
  Corrections to Leptogenesis}, \emph{Nucl. Phys.} {\bf B838}, \penalty0 1--27,
   (2010).

\bibitem{Anisimov:2010dk}
A.~Anisimov, W.~{Buchm\"uller}, M.~Drewes, and S.~Mendizabal, {Quantum
  Leptogenesis I}, \emph{Annals Phys.} {\bf 326}, \penalty0 1998--2038,
  (2011).
\newblock [Erratum: Annals Phys.338,376(2011)].

\bibitem{Garny:2011hg}
M.~Garny, A.~Kartavtsev, and A.~Hohenegger, {Leptogenesis from first principles
  in the resonant regime}, \emph{Annals Phys.} {\bf 328}, \penalty0 26--63,
  (2013).

\bibitem{Frossard:2012pc}
T.~Frossard, M.~Garny, A.~Hohenegger, A.~Kartavtsev, and D.~Mitrouskas,
  {Systematic approach to thermal leptogenesis}, \emph{Phys. Rev.} {\bf
  D87}\penalty0 (8), \penalty0 085009,  (2013).

\bibitem{Garny:2010nj}
M.~Garny, A.~Hohenegger, and A.~Kartavtsev, {Medium corrections to the
  CP-violating parameter in leptogenesis}, \emph{Phys. Rev.} {\bf D81},
  \penalty0 085028,  (2010).

\bibitem{Garbrecht:2010sz}
B.~Garbrecht, {Leptogenesis: The Other Cuts}, \emph{Nucl. Phys.} {\bf B847},
  \penalty0 350--366,  (2011).

\bibitem{Garbrecht:2013iga}
B.~Garbrecht and M.~J. Ramsey-Musolf, {Cuts, Cancellations and the Closed Time
  Path: The Soft Leptogenesis Example}, \emph{Nucl. Phys.} {\bf B882},
  \penalty0 145--170,  (2014).

\bibitem{Dev:2014laa}
P.~S. Bhupal~Dev, P.~Millington, A.~Pilaftsis, and D.~Teresi, {Flavour
  Covariant Transport Equations: an Application to Resonant Leptogenesis},
  \emph{Nucl. Phys.} {\bf B886}, \penalty0 569--664,  (2014).

\bibitem{Baym:1961zz}
G.~Baym and L.~P. Kadanoff, {Conservation Laws and Correlation Functions},
  \emph{Phys. Rev.} {\bf 124}, \penalty0 287--299,  (1961).

\bibitem{Keldysh:1964ud}
L.~V. Keldysh, {Diagram technique for nonequilibrium processes}, \emph{Zh.
  Eksp. Teor. Fiz.} {\bf 47}, \penalty0 1515--1527,  (1964).
\newblock [Sov. Phys. JETP20,1018(1965)].

\bibitem{Drewes:2012qw}
M.~Drewes, S.~Mendizabal, and C.~Weniger, {The Boltzmann Equation from Quantum
  Field Theory}, \emph{Phys. Lett.} {\bf B718}, \penalty0 1119--1124,  (2013).

\bibitem{Berges:2002wt}
J.~Berges and M.~M. Muller.
\newblock {Nonequilibrium quantum fields with large fluctuations}.
\newblock In \emph{{285th Heraeus Seminar: Interdisciplinary Workshop on
  Progress in Nonequilibrium Greens Functions (Kadanoff-Baym Equations II)
  Dresden, Gremany, August 19-23, 2002}},  (2002).
\newblock URL \url{http://alice.cern.ch/format/showfull?sysnb=2337540}.

\bibitem{Prokopec:2003pj}
T.~Prokopec, M.~G. Schmidt, and S.~Weinstock, {Transport equations for chiral
  fermions to order h bar and electroweak baryogenesis. Part 1}, \emph{Annals
  Phys.} {\bf 314}, \penalty0 208--265,  (2004).

\bibitem{Garny:2010nz}
M.~Garny, A.~Hohenegger, and A.~Kartavtsev, {Quantum corrections to
  leptogenesis from the gradient expansion}, {\ttfamily arXiv:1005.5385},
  (2010).

\bibitem{Wentzel:1926aor}
G.~Wentzel, {Eine Verallgemeinerung der Quantenbedingungen für die Zwecke der
  Wellenmechanik}, \emph{Z. Phys.} {\bf 38}\penalty0 (6), \penalty0 518--529,
  (1926).

\bibitem{Kramers:1926njj}
H.~A. Kramers, {Wellenmechanik und halbzahlige Quantisierung}, \emph{Z. Phys.}
  {\bf 39}\penalty0 (10), \penalty0 828--840,  (1926).

\bibitem{Brillouin:1926blg}
L.~Brillouin, {La mécanique ondulatoire de Schrödinger; une méthode
  générale de resolution par approximations successives}, \emph{Compt. Rend.
  Hebd. Seances Acad. Sci.} {\bf 183}\penalty0 (1), \penalty0 24--26,  (1926).

\bibitem{Landau:1980mil}
L.~D. Landau and E.~M. Lifshitz, \emph{{Statistical Physics, Part 1}}. vol.~5,
  \emph{Course of Theoretical Physics}, (Butterworth-Heinemann, Oxford, 1980).
\newblock ISBN 9780750633727.

\bibitem{Asaka:2006rw}
T.~Asaka, M.~Laine, and M.~Shaposhnikov, {On the hadronic contribution to
  sterile neutrino production}, \emph{JHEP}. {\bf 06}, \penalty0 053,  (2006).

\end{thebibliography}

\end{document}